\documentclass[prb,floatfix,twocolumn,showpacs,amsmath,amssymb,reprint]{revtex4-1}
\usepackage{graphicx}
\usepackage{dcolumn}
\usepackage{bm}
\begin{document}


\title{
Merging Dirac points and topological phase transitions 
in the tight-binding model \\
on the generalized honeycomb lattice
}

\author{Yasumasa Hasegawa}
\affiliation{Department of Material Science, 
Graduate School of Material Science, 
University of Hyogo, Hyogo, 678-1297, Japan}

\author
{Keita Kishigi}
\affiliation{Faculty of Education, 
Kumamoto University, Kurokami 2-40-1, Kumamoto, 860-8555, Japan}

\date{\today}

\begin{abstract}
Moving, merging and annihilating Dirac points are studied theoretically
in the tight-binding model on honeycomb lattice 
with up-to third-nearest-neighbor
hoppings.
We obtain a rich phase diagram 
of the topological phase transitions in the parameter space of 
direction-dependent hoppings.
We obtain the conditions for the three Dirac points to merge
and for the tricritical points.
We find that only very small third-nearest-neighbor hoppings
are enough for the 
existence of the merging of three-Dirac-points 
 and the tricritical points, if the system is sufficiently anisotropic.
 The density of states is obtained to be
$D(\epsilon) \propto |\epsilon|^{1/3}$ when three Dirac points merge, and 
$D(\epsilon) \propto |\epsilon|^{1/4}$ at the tricritical points.
It is possible to realize these  topological phase transitions 
in the ultracold atoms on the optical lattice,
strained monolayer graphene or strained bilayer graphene.
\end{abstract}

\date{July 30, 2012}

\pacs{
73.22.Pr, 73.43.Nq, 71.10.Pm, 73.20.At}
\maketitle

\section{Introduction}
Massless Dirac fermions are realized in various 
fields in condensed matter physics and they attract much interest recently, 
e.g.  graphene
\cite{Novoselov2004,Novoselov2005,Zhang2005}, 
the organic conductor $\alpha$-(BEDT-TTF)$_2$I$_3$%
\cite{Katayama2006,Tajima2009} 
and iron-based superconductor, BaFe$_2$As$_2$
\cite{Richard2010}. 
These materials have Dirac points and the linear band structure 
near the Fermi energy.
%

The moving, merging and annihilating Dirac points
due to the breaking of the rotational symmetry
were studied\cite{Hasegawa2006} and
observed recently in the 
ultracold atoms on a optical lattice\cite{Tarruel2012}.
Many authors have  theoretically studied the merging Dirac points
in $\alpha$-(BEDT-TTF)$_2$I$_3$\cite{Montambaux2009},
strained or twisted 
bilayer graphene\cite{Son2011,Mucha2011,Gail2011,Gail2012,Montambaux2012},
and the honeycomb lattice 
with third-nearest-neighbor hoppings\cite{Bena2011}.
Most of the studies have used the effective model,
which describe the energy near the Dirac points.

The energy band of graphene is obtained by first-principles 
band calculations\cite{Farjam2009,Gui2009,Choi2010},
and the band structure near the Fermi energy can be reproduced
by a simple tight-binding model  
\cite{wallace}.
If hoppings are only between the nearest sites 
and they are independent of the direction, 
there are two Dirac points at 
$\mathbf{K}$ and $\mathbf{K}'$ in the Brillouin zone. 
Reich {\it et al.}\cite{Reich2002} 
have shown that  up to third-nearest-neighbor 
hoppings are necessary to describe the band structure in all Brillouin zone.
They have obtained  that the nearest-neighbor, 
the next-nearest-neighbor and the third-nearest-neighbor hoppings 
are 2.79-2.97 eV, 0.073-0.68eV and 0.3-0.33 eV, 
respectively\cite{Reich2002}. 

A finite gap is important for the application to nano devices.
There are two routes to open the finite gap at
the Dirac points. One route is via the the violation of the 
inversion symmetry. The violation of the inversion symmetry
can be caused by the difference of the on-site 
potentials on A and B sublattices or the sublattice-dependent second-neighbor
hoppings\cite{Kishigi2008,Kishigi2008b}.
The finite gap observed experimentally in graphene 
on a SiC substrate\cite{zou2007} 
is caused by the breaking of the inversion symmetry. 

The other route to open a finite gap 
is realized by making two Dirac points with opposite 
topological number merge
and annihilate\cite{Hasegawa2006,Montambaux2009,%
Son2011,Mucha2011,Gail2011,Gail2012,Montambaux2012,Bena2011}. 
When the hoppings between nearest 
sites become different in three directions,
the Dirac points move from $\mathbf{K}$ and $\mathbf{K}'$  
to one of the 
three inequivalent ${M}$ points ($\mathbf{M}_1$, $\mathbf{M}_2$ 
and $\mathbf{M}_3$),
merge into a semi-Dirac point
and annihilate to make a finite energy gap. 
In order to make a finite gap in monolayer graphene, 
we need the deformations 
of the order of 20\% caused by the strong uniaxial 
strain \cite{Pereira2009} or the shear strain\cite{Cocco2010}, 
which are difficult to realize. 
However, the parameters can be controlled in the ultracold
atoms on the optical lattice and the merging and annihilating Dirac points
are observed\cite{Tarruel2012}.

It has been shown that the strained or twisted 
bilayer graphene is modeled by the single-layer honeycomb lattice with
the third-nearest-neighbor hoppings, the strength of which are near the half  
of the nearest-neighbor hoppings\cite{Montambaux2012}.

In this paper we study the tight-binding model on the
honeycomb lattice with up-to
third-nearest-neighbor hoppings, and we study the moving and the merging
of the Dirac points by changing the 
strength of direction-dependent hoppings.
We obtain interesting phase diagrams, in which we can see
 the merging of three Dirac points and the tricritical points.
The density of states due to Dirac points 
and merged Dirac points are calculated.

The model is given in Section~\ref{section2}. 
The topological number and the Berry phase of the Dirac points 
are discussed in Section~\ref{section3}.
In Section~\ref{section4} we study the phase diagram
in which we consider only the phase transition
due to the merging of two Dirac points at 
$\mathbf{M}_1$,  $\mathbf{M}_2$, $\mathbf{M}_3$, and $\boldsymbol{\Gamma}$.
Moving Dirac points are studied in the model with 
direction-dependent nearest-neighbor hoppings 
in Section \ref{section5}.
Rich phase diagrams in the model with 
direction-dependent nearest-neighbor and third-nearest-neighbor hoppings
are studied in Section~\ref{section6}.
Section~\ref{conclusions} is the conclusions and
detailed calculations are given in Appendix.

\section{Tight-binding model on honeycomb lattice}
\label{section2}
\begin{figure}[tb]
\begin{center}
\includegraphics[width=0.2\textwidth]{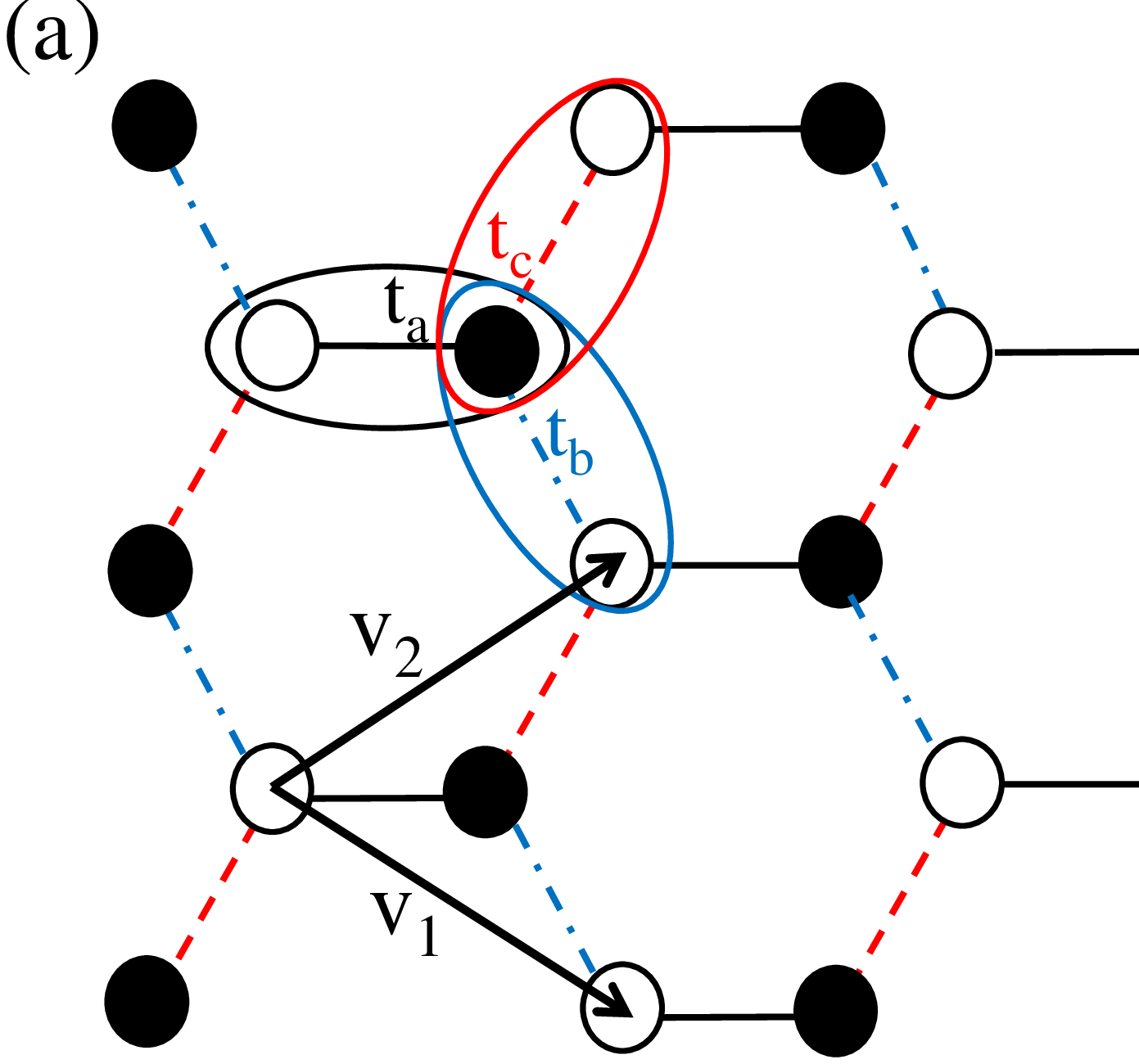}\\
\vspace{1cm}
\includegraphics[width=0.2\textwidth]{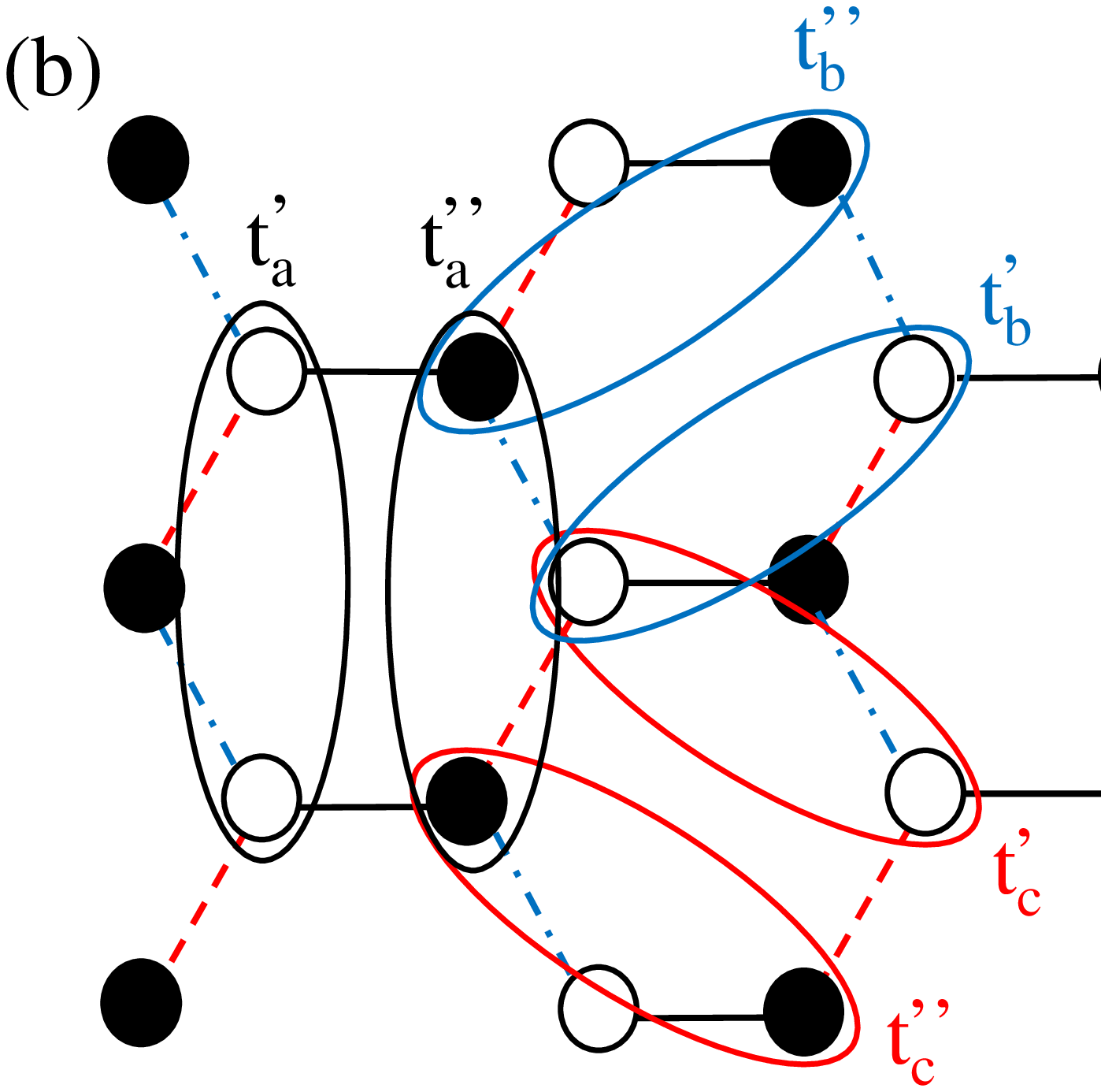}\\
\vspace{1cm}
\includegraphics[width=0.2\textwidth]{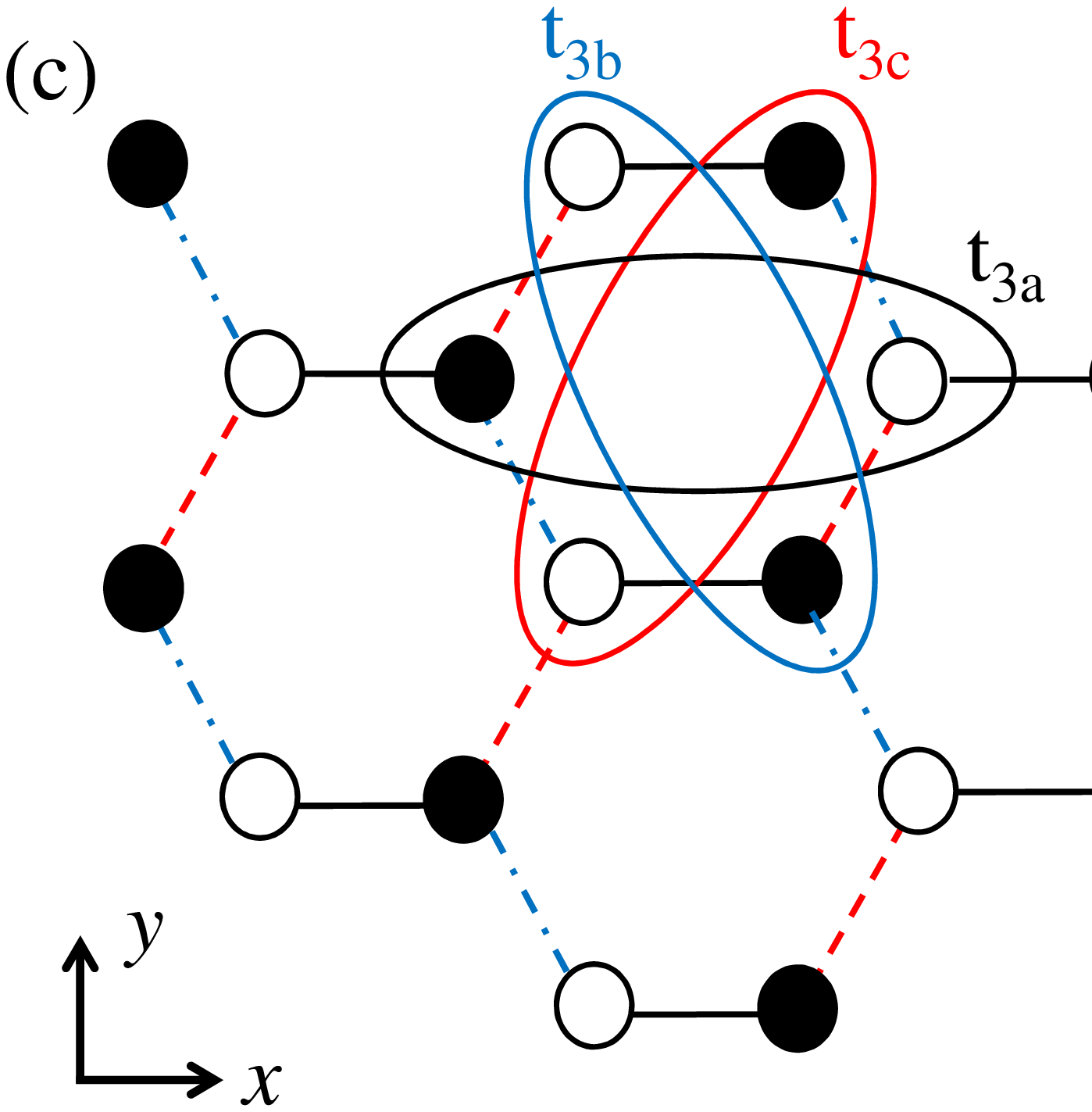}\\
\end{center}
\caption{(Color online)
Honeycomb lattice and hoppings. Filled circles ($\bullet$)
 and open circles ($\circ$) are
sites in  A sublattice 
and B sublattice, respectively.
$v_1$ and $v_2$ are the unit vectors. 
The hoppings between the nearest sites, the next-nearest sites,
and the third-nearest sites are shown in (a), (b), and (c), respectively,
and they are considered to depend on the direction and the sublattices. 
}
\label{figlattice}
\end{figure}
We study the tight-binding electrons on the
honeycomb lattice (see Fig.~\ref{figlattice}).
There are two sublattices, A and B
(open and filled circles in Fig.~\ref{figlattice}), 
which form triangular lattices.
We take the unit vectors as 
\begin{equation}
\mathbf{v}_1 = \left( \frac{\sqrt{3}}{2} a, -\frac{1}{2} a\right),
\end{equation}
and
\begin{equation}
\mathbf{v}_2 = \left( \frac{\sqrt{3}}{2} a,  \frac{1}{2} a\right),
\end{equation}
and the vectors connecting the nearest sites as
\begin{align}
 \boldsymbol{\delta}_a &= \frac{1}{3} (\mathbf{v}_1+\mathbf{v}_2)
 = (\frac{\sqrt{3}}{3}a,0), \\
 \boldsymbol{\delta}_b &= \frac{1}{3} (\mathbf{v}_2-2\mathbf{v}_1)
 =(- \frac{\sqrt{3}}{6}a, \frac{1}{2}a), 
\end{align}
and
\begin{equation}
 \boldsymbol{\delta}_c = \frac{1}{3} (\mathbf{v}_1-2\mathbf{v}_2)
  =(- \frac{\sqrt{3}}{6}a, -\frac{1}{2}a),
\end{equation}
where $a$ is the lattice constant. Hereafter, we set $a=1$. 

We consider the direction-dependent hoppings between nearest sites
($t_a$, $t_b$, and $t_c$), 
 the next-nearest sites ($t_a'$, $t_b'$, $t_c'$,
$t_a''$, $t_b''$ and $t_c''$)
and the third-nearest sites ($t_{3a}$, $t_{3b}$, and $t_{3c}$), 
as shown in Fig.~\ref{figlattice}.
The hoppings between the next-nearest sites within the 
A sublattice are labeled by $t_a'$, $t_b'$ and $t_c'$, and 
those within the B sublattice are labeled by $t_a''$, $t_b''$ and $t_c''$.
The Hamiltonian is given by
\begin{equation}
 \mathcal{H} = \sum_{\mathbf{k}} c_{\mathbf{k}}^{\dagger} 
\mathcal{E}_{\mathbf{k}} c_{\mathbf{k}},
\end{equation}
where
\begin{equation}
 c_{\mathbf{k}} = \left(
\begin{array}{c}
 a_{\mathbf{k}} \\
 b_{\mathbf{k}}
\end{array} \right),
\end{equation}
$a_{\mathbf{k}}$ and $b_{\mathbf{k}}$ are the
annihilation operators at $A$ and $B$ sublattices, respectively, 
\begin{equation}
\mathcal{E}_{\mathbf{k}} = \sum_{\mu=0,1,2,3} \epsilon_{\mu}(\mathbf{k})
 \sigma_{\mu},
\label{eqmathcalE}
\end{equation}
$\sigma_0$ is the $2 \times 2$ unit matrix, $\sigma_j$ ($j=1$, $2$, and $3$)
are Pauli matrices and
\begin{align}
\epsilon_0(\mathbf{k})
=&\frac{1}{2} (\epsilon_A +\epsilon_B) 
-(t_c' + t_c'') \cos (\mathbf{v}_1 \cdot \mathbf{k}) 
\nonumber \\
&-(t_b' + t_b'') \cos (\mathbf{v}_2 \cdot \mathbf{k})
\nonumber \\
&-(t_a' + t_a'') \cos ((\mathbf{v}_2-\mathbf{v}_1) 
\cdot \mathbf{k}), 
\label{eqepsilon0}
\end{align}
\begin{align}
\epsilon_1(\mathbf{k})
=& 
-t_a \cos ( \boldsymbol{\delta}_a \cdot \mathbf{k}) 
-t_b \cos ( \boldsymbol{\delta}_b \cdot \mathbf{k})
\nonumber \\
&-t_c \cos (\boldsymbol{\delta}_c \cdot \mathbf{k})\nonumber \\
&-t_{3a} \cos (2\boldsymbol{\delta}_a \cdot \mathbf{k}) 
-t_{3b}  \cos (2\boldsymbol{\delta}_b \cdot \mathbf{k})
\nonumber \\
&-t_{3c} \cos (2\boldsymbol{\delta}_c \cdot \mathbf{k}),
\label{eqepsilon1}
\\
\epsilon_2(\mathbf{k})
=& 
 -t_a \sin (\boldsymbol{\delta}_a \cdot \mathbf{k}) 
 -t_b \sin (\boldsymbol{\delta}_b \cdot \mathbf{k})
\nonumber \\
&-t_c \sin (\boldsymbol{\delta}_c \cdot \mathbf{k}) \nonumber \\
&-t_{3a} \sin (-2\boldsymbol{\delta}_a \cdot \mathbf{k}) 
 -t_{3b} \sin (-2\boldsymbol{\delta}_b \cdot \mathbf{k})
  \nonumber \\
&-t_{3c} \sin (-2\boldsymbol{\delta}_c \cdot \mathbf{k}),
\label{eqepsilon2} \\
\epsilon_3(\mathbf{k})
=&\frac{1}{2} (\epsilon_A - \epsilon_B) 
-(t_c' - t_c'') \cos (\mathbf{v}_1 \cdot \mathbf{k}) 
\nonumber \\
&-(t_b' - t_b'') \cos (\mathbf{v}_2 \cdot \mathbf{k})
\nonumber \\
&-(t_a' - t_a'') \cos ((\mathbf{v}_2-\mathbf{v}_1) \cdot \mathbf{k}).
\label{eqepsilon3}
\end{align}
The energy is given by
\begin{equation}
E_{\pm}(\mathbf{k}) = \epsilon_0(\mathbf{k}) \pm 
\sqrt{\sum_{j=1}^3\left(\epsilon_j(\mathbf{k})\right)^2}.
\end{equation}

The nearest-neighbor hoppings and the third-nearest-neighbor hoppings
connect the sites in A and B sublattices, 
while the next-nearest hoppings connect 
sites within A sublattice or B sublattice.
Therefore, the nearest hoppings and the third-nearest-neighbor hoppings
appear only in the off-diagonal elements (Eqs.~(\ref{eqepsilon1})
and  (\ref{eqepsilon2})), and the next-nearest 
hoppings appear only in the diagonal elements (Eqs.~(\ref{eqepsilon0}) 
and (\ref{eqepsilon3})).

There are no gaps at $\mathbf{k}=\mathbf{k}^*$,
when $\epsilon_1(\mathbf{k}^*)=\epsilon_2(\mathbf{k}^*)
=\epsilon_3(\mathbf{k}^*)=0$.
In general, 
each of the equations $\epsilon_i(\mathbf{k})=0$ 
($i=1$, $2$, and $3$) gives the
line in the plane of $k_x$ and $k_y$. 
Therefore, it is only in rare case that three lines 
($\epsilon_i(\mathbf{k})=0$ ($i=1$, $2$, and $3$)) 
intersect at the same point\cite{Kishigi2008,Kishigi2008b}.
However, when the system has an inversion symmetry, i.e.,
A and B sublattices are equivalent,
we get
\begin{equation}
\epsilon_3(\mathbf{k})=0,
\label{eqe30}
\end{equation}
and there exist  Dirac points where two lines
($\epsilon_1(\mathbf{k})=0$ and $\epsilon_2(\mathbf{k})=0$)
intersect.
Hereafter, we study the system with inversion symmetry, i.e., 
Eq.~(\ref{eqe30}) is satisfied.
 
As we have discussed, 
the next-nearest-neighbor
hoppings appear only in $\epsilon_0(\mathbf{k})$
and $\epsilon_3(\mathbf{k})$. 
Therefore, the next-nearest-neighbor
hoppings do not affect the existence and 
the location of the Dirac points in the system with inversion symmetry.
They only affect the energy dispersion near the Dirac points, i.e. 
they may tilt a Dirac cone.
Since we are interested in the location of the Dirac 
points and the topological 
phase transitions, we ignore
the effects of the next-nearest-neighbor hoppings
in this paper.

\section{topological number}
\label{section3}
We define $\epsilon(\mathbf{k})$ and $\phi(\mathbf{k})$ by
\begin{align}
 \epsilon_1(\mathbf{k}) &=\epsilon(\mathbf{k}) \cos ( \phi(\mathbf{k})),
\label{eqdefphi1} \\
 \epsilon_2(\mathbf{k}) &=\epsilon(\mathbf{k}) \sin ( \phi(\mathbf{k})).
\label{eqdefphi2}
\end{align}
Then the matrix given in Eq.~(\ref{eqmathcalE}) is
written as
\begin{equation}
 \mathcal{E}_{\mathbf{k}} = \epsilon(\mathbf{k})
 \left(  \cos (\phi(\mathbf{k})) \sigma_1
       + \sin (\phi(\mathbf{k})) \sigma_2 \right),
\end{equation} 
and it is diagonalized as
\begin{equation}
  U_{\mathbf{k}}^{-1} \mathcal{E}_{\mathbf{k}} U_{\mathbf{k}} 
= \epsilon(\mathbf{k}) \sigma_3,
\end{equation}
where
\begin{align}
 U_{\mathbf{k}} &= 
 e^{-i \frac{\phi(\mathbf{k})}{2} \sigma_3} 
e^{-i \frac{\pi}{4} \sigma_2} \nonumber \\
 &=\frac{1}{\sqrt{2}} \left( \begin{array}{cc}
 e^{-i \frac{\phi(\mathbf{k})}{2}} & -e^{-i \frac{\phi(\mathbf{k})}{2}} \\
 e^{ i \frac{\phi(\mathbf{k})}{2}} &  e^{ i \frac{\phi(\mathbf{k})}{2}}
\end{array} \right).
\end{align}
Therefore,
the vectors,
\begin{equation}
  \Psi_{\pm} (\mathbf{k}) = \frac{1}{\sqrt{2}} 
 \left(  \begin{array}{c}
  \pm e^{-i \frac{\phi(\mathbf{k})}{2}}   \\
      e^{ i \frac{\phi(\mathbf{k})}{2}}  
\end{array} \right),
\label{eqeigenvec}
\end{equation}
are the eigenvectors of the matrix  $\mathcal{E}_{\mathbf{k}}$
with eigenvalues $\pm \epsilon(\mathbf{k})$.
%
\begin{figure}[bt]
\begin{flushleft}\hspace*{0.5cm} (a)\end{flushleft}
\begin{center}
\includegraphics[width=0.29\textwidth]{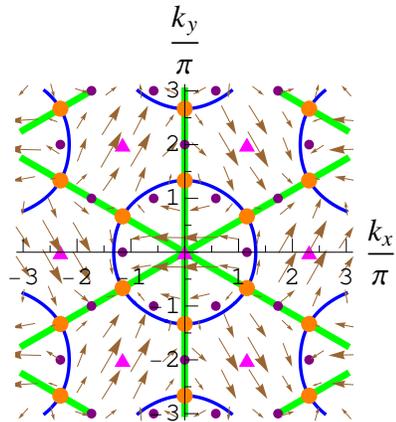}
\end{center}
\begin{flushleft}\hspace*{0.5cm} (b)\end{flushleft}
\begin{center}
\includegraphics[width=0.29\textwidth]{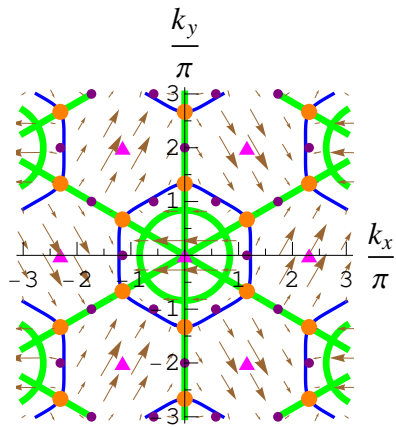}
\end{center}
\caption{(Color online)
Lines for $\epsilon_1(\mathbf{k})=0$ (thin blue lines)
and  $\epsilon_2(\mathbf{k})=0$ (thick green lines).
Parameters are 
 $t_a=t_b=t_c=1$, $t_{3a}=t_{3b}=t_{3c}=0$ in (a),
and  $t_a=t_b=t_c=1$, $t_{3a}=t_{3b}=t_{3c}=0.2$ in (b).
Orange circles show $\mathbf{K}$, $\mathbf{K}'$ and their equivalent points.
Small purple circles 
are $\mathbf{M}_i$ ($i=1,2$ and $3$)
and their equivalent points,
and  triangles are  $\boldsymbol{\Gamma}$ and their equivalent points.
Arrows show the vectors, $(\epsilon_1(\mathbf{k}), \epsilon_2(\mathbf{k}))$.
Dirac points appear at the intersection points of the blue lines and 
the green lines,
which are $\mathbf{K}$ and $\mathbf{K}'$ 
in both figures.
New green circular lines appear around 
$\mathbf{k}=0$, if $t_{3a} > t_a/8$ as shown in (b). 
The Dirac points are, however,
only at the $\mathbf{K}$ and $\mathbf{K}'$, if $t_{3a} < t_a/3$. 
The topological numbers are
 $+1$ and $-1$ for $\mathbf{K}$ and $\mathbf{K}'$, respectively.
}
\label{figmath1}
\end{figure}
%

In Fig.~\ref{figmath1}, we plot the lines given by
$\epsilon_1(\mathbf{k})=0$ (thin blue lines)
and $\epsilon_2(\mathbf{k})=0$ (thick green lines)
for $t_a=t_b=t_c=1$,  $t_{3a}=t_{3b}=t_{3c}=0$ in (a) 
and  $t_a=t_b=t_c=1$, $t_{3a}=t_{3b}=t_{3c}=0.2$ in (b).
The Dirac points are given by the intersection points
of the blue lines and the green lines, which are at 
\begin{equation}
 \mathbf{K} = (0, \frac{4 \pi}{3}),
\end{equation}
\begin{equation}
 \mathbf{K}' = (0, - \frac{4 \pi}{3}),
\end{equation}
and their equivalent points
related by the reciprocal lattice vectors.
We also plot the vector
$(\epsilon_1(\mathbf{k}), \epsilon_2(\mathbf{k}))$
in Fig.~\ref{figmath1}. The Dirac points are the cores of the vortex
for the vector $(\epsilon_1(\mathbf{k}), \epsilon_2(\mathbf{k}))$.
The topological number is $+1$ and $-1$ 
at $\mathbf{K}$ and $\mathbf{K}'$, respectively.
Note that the configurations of the arrows
have a topological number $+1$, if they 
are clockwisely or counterclockwisely rotating 
flows around the core, 
or if all arrows are pointing out  
or into the cores,
since $\phi(\mathbf{k})$ changes $+2 \pi$ when $\mathbf{k}$ goes 
counterclockwisely around the core.
The configurations of the arrows
have a topological number $-1$, if they are pointing out the core 
in $\pm x$ direction and pointing into the core
 in the $\pm y$ direction, for example.

Since the eigenstates are given by $\phi(\mathbf{k})$
(Eq.~(\ref{eqeigenvec})), the topological number $\pm 1$ 
is related to the Berry phase of $\pm \pi$. 

\section{merging Dirac points at $\mathbf{M}$  or $\boldsymbol{\Gamma}$ 
in the Brillouin zone}
\label{section4}
\begin{figure}[bt]
\flushleft{\hspace*{0.5cm} (a) \hfill}  \\
\begin{center}
%
\includegraphics[width=0.35\textwidth]{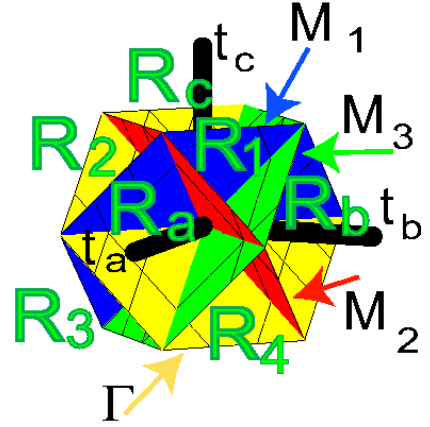}  \\
\end{center}
\flushleft{\hspace*{0.5cm} (b) } \\
\begin{center}
%
\includegraphics[width=0.35\textwidth]{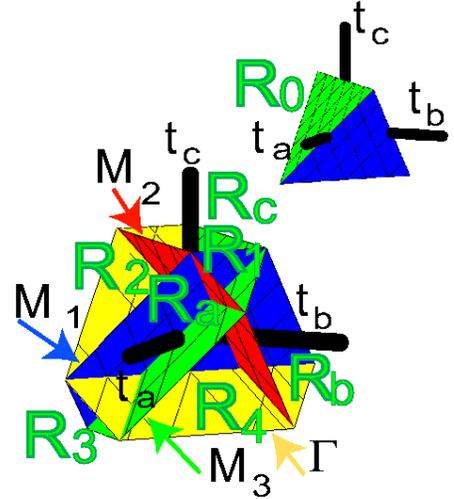}
\end{center}
\caption{(Color online)
Planes given by Eqs. ~(\ref{eqm10}) - (\ref{eqm00})
in the parameter space of $t_a$, $t_b$ and $t_c$
in the case of
 (a) $t_{3a}=t_{3b}=t_{3c} =0$
and (b) $t_{3a}=t_{3b}=t_{3c} >0$.
On the planes labeled by $M_1$, $M_2$, $M_3$ and $\Gamma$,
two Dirac points merge at  
$\mathbf{M}_1$,
$\mathbf{M}_2$, 
$\mathbf{M}_3$, 
and $\boldsymbol{\Gamma}$, respectively.
The 3D space is divided into 14 regions, $R_1$ - $R_4$, $R_a$, $R_b$, $R_c$, 
 $R_1^*$ - $R_4^*$, $R_a^*$, $R_b^*$, and $R_c^*$
when $t_{3a}=t_{3b}=t_{3c} =0$. The regions,
 $R_1^*$, $R_2^*$ etc. 
are given by the inversion of  $R_1$, $R_2$ etc..
When  $t_{3a}=t_{3b}=t_{3c} \neq 0$,
a regular tetrahedral region $R_0$ appears around the origin,
as shown in the inset of (b).
Topological phase transitions occur and the number of the Dirac points 
changes by two when parameters intersect the planes, 
$M_1$, $M_2$, $M_3$ and $\Gamma$.}
\label{figparam013d}
\end{figure}
\begin{figure}[bt]
\begin{center}
\includegraphics[width=0.3\textwidth]{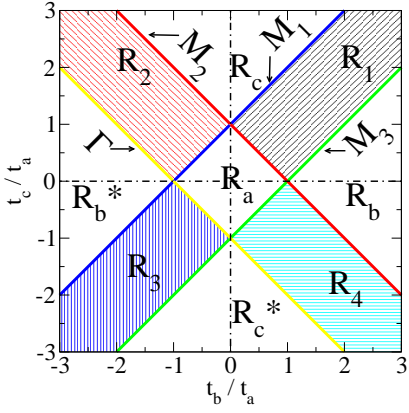} 
\end{center}
\caption{(Color online)
Phase diagram in the plane of $t_b/t_a$-$t_c/t_a$ for 
the tight-binding electrons on
honeycomb lattice with direction-dependent nearest-neighbor hoppings
($t_{3a}=t_{3b}=t_{3c}=0$).
}
\label{figpara2d}
\end{figure}
\begin{figure}[bt]
%
\begin{flushleft} \hspace{0.5cm}  (a) \end{flushleft} 
\begin{center}
\includegraphics[width=0.35\textwidth]{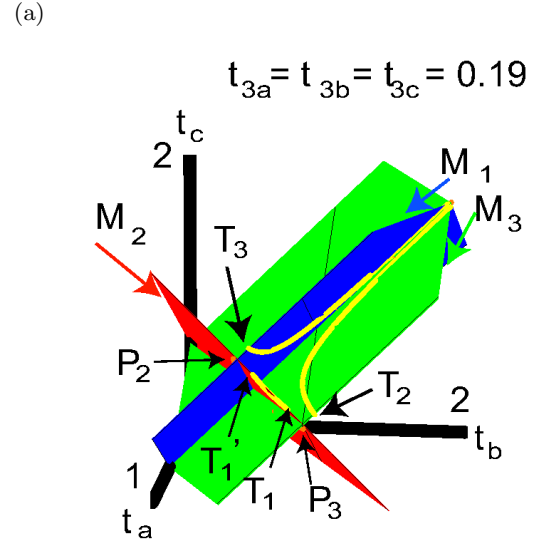}
\end{center}
\begin{flushleft} \hspace{0.5cm}  (b) \end{flushleft} 
\begin{center}
\includegraphics[width=0.35\textwidth]{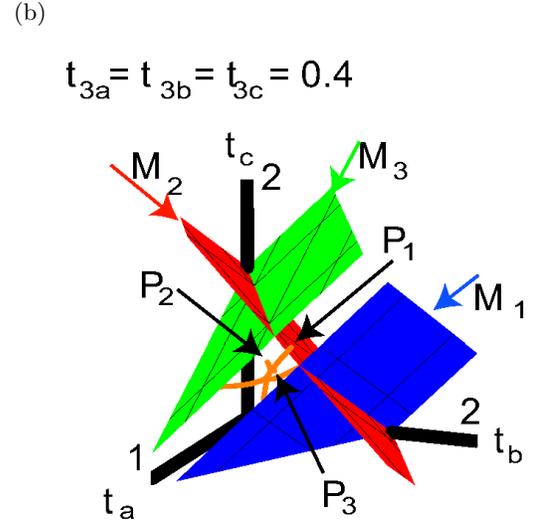}
\end{center}
\caption{(Color online)
3D plot of the phase diagrams for (a) $t_{3a}=t_{3b}=t_{3c}=0.19$
 and (b) $0.4$ .
The intersection with the plane $t_a=1$
is the phase diagram in Fig.~\ref{figparam4x} 
and Fig.~\ref{fig16_gx}~(b).
Three Dirac points merge when parameters are at $\mathbf{P}_1$,
$\mathbf{P}_2$, or $\mathbf{P}_3$,
and $\mathbf{T}_1$, $\mathbf{T}_1'$, $\mathbf{T}_2$, and $\mathbf{T}_3$
are tricritical points, which are discussed in Section~\ref{section6} 
and Appendix. }
\label{figfig6}
\end{figure}
%
When
$\mathbf{k}^*$ is a Dirac points, $-\mathbf{k}^*$ 
is also a Dirac point if the system is invariant with respect 
to the time-reversal operation, which is always the case in this paper. 
Two Dirac points, which are exchanged by time-reversal operation,
can merge only at the points $\mathbf{k}^*$ where $\mathbf{k}^*$ and 
$-\mathbf{k}^*$
are the same or are different by the reciprocal lattice vectors.
These points are
only four points
in the Brillouin zone, $\mathbf{M}_1$, $\mathbf{M}_2$, $\mathbf{M}_3$, 
and $\boldsymbol{\Gamma}$, 
\begin{align}
\mathbf{M}_1&=(\frac{\sqrt{3}\pi}{3},\pi), \\
\mathbf{M}_2&=(\frac{2\sqrt{3}\pi}{3}, 0), \\
\mathbf{M}_3&=(\frac{\sqrt{3}\pi}{3},-\pi), \\
\boldsymbol{\Gamma}&=(0,0).
\end{align}

 From the equation
\begin{equation}
\epsilon_1(\mathbf{M}_1) =\epsilon_2(\mathbf{M}_1)=0,
\end{equation}
we obtain the condition for two
Dirac points to merge at 
$\mathbf{M}_1$ as
\begin{equation}
  t_a + t_b - t_c =t_{3a}+t_{3b}+t_{3c}  \hspace{1cm}: \mathbf{M}_1.
\label{eqm10}
\end{equation}
Similarly we obtain the conditions for two
Dirac points to merge at 
$\mathbf{M}_2$, $\mathbf{M}_3$, and $\boldsymbol{\Gamma}$  as
\begin{align}
 -t_a + t_b + t_c =t_{3a}+t_{3b}+t_{3c} & \hspace{1cm}: \mathbf{M}_2 ,
\label{eqm20}\\
  t_a - t_b + t_c =t_{3a}+t_{3b}+t_{3c} & \hspace{1cm}: \mathbf{M}_3 ,
\label{eqm30}
\end{align}
and
\begin{align}
 -t_a - t_b - t_c =t_{3a}+t_{3b}+t_{3c} & \hspace{1cm}: \boldsymbol{\Gamma},
\label{eqm00}
\end{align}
respectively.
We plot the planes given by Eqs.~(\ref{eqm10}) - (\ref{eqm00})
in the parameter space in $t_a$, $t_b$ and $t_c$
in the cases of $t_{3a}=t_{3b}=t_{3c}=0$ and $t_{3a}=t_{3b}=t_{3c}>0$
in Fig.~\ref{figparam013d}.
The phase diagram in $t_b/t_a - t_c/t_a$ plane 
(Fig.~\ref{figpara2d}) can be
obtained as the intersection of the 
phase diagram in 3D plot (Fig.~\ref{figparam013d}) and the plane
of constant $t_a$, as shown in Fig.~\ref{figfig6}.

If $t_{3a}+t_{3b}+t_{3c}=0$,
the $t_a-t_b-t_c$ space
is divided into 14 regions by four planes, 
Eqs.~(\ref{eqm10}) - (\ref{eqm00}),
as shown in Fig.~\ref{figparam013d} (a).
 There are two gapless Dirac points when parameters are 
in the regions labeled by $R_1$, $R_2$, $R_3$ and $R_4$ 
 and their inversion regions, 
$R_1^*$, $R_2^*$, $R_3^*$ and $R_4^*$ in Fig.~\ref{figparam013d}. 
The regions, $R_1$, $R_2$, $R_3$ and $R_4$, 
correspond to the shaded regions in Fig.~\ref{figpara2d},
where two-dimensional plot of the phase diagram is 
shown in $t_b/t_a$ - $t_c/t_a$ plane.
On the planes and lines ($M_1$, $M_2$, $M_3$ and $\Gamma$)
in Fig.~\ref{figparam013d} and Fig.~\ref{figpara2d},
topological phase transition occurs and
 two Dirac points merges at $\mathbf{M}_1$, $\mathbf{M}_2$, $\mathbf{M}_3$ 
or $\boldsymbol{\Gamma}$ in the Brillouin zone.  
The phase diagram for $t_{3a}+t_{3b}+t_{3c}=0$ is the same as obtained 
previously\cite{Hasegawa2006}.

We show the 3D plot of the phase diagram in Fig.~\ref{figparam013d}~(b) 
for $t_{3a}+t_{3b}+t_{3c} > 0$. Although it is invisible
in the main figure in Fig.~\ref{figparam013d}~(b),  
there is a regular tetrahedron region
$R_0$ around the origin $(0,0,0)$ as shown in the inset
in Fig.~\ref{figparam013d}~(b).
%
%
\begin{figure}[bt]
\begin{center}
\includegraphics[width=0.4\textwidth]{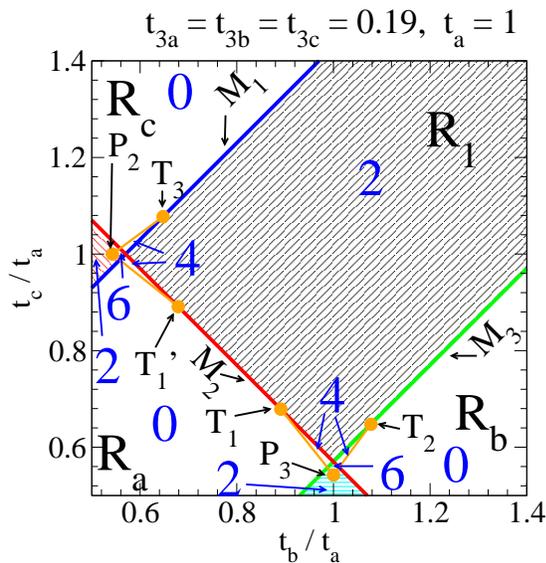}
%
\end{center}
\caption{(Color online)
Phase diagrams in the $t_b/t_a$ - $t_c/t_a$ plane 
for $t_a=1$ and $t_{3a}=t_{3b}=t_{3c}=0.19$.
When the third-nearest-neighbor hoppings are finite,
phase transition lines $M_1$, $M_2$, $M_3$ and $\Gamma$ are shifted
and new transition lines appear, as discussed in Section \ref{section6}
and Appendix. 
The phase diagrams for other values of third-nearest-neighbor hoppings 
calculated numerically
are show in Fig.~\ref{figphasenum} and Fig.~\ref{fig16_gx}.
The numbers in the figures indicate the numbers of the Dirac points
in the Brillouin zone.
}
\label{figparam4x}
\end{figure}
%
%
%
We plot the phase diagram in the 
plane of $t_b/t_a$ - $t_c/t_a$ plane for $t_a=1$ and 
$t_{3a}=t_{3b}=t_{3c}=0.19$
in Fig.~\ref{figparam4x}.
If the third-nearest-neighbor hoppings are finite, 
the number of the Dirac points can be larger than 2
and 
Dirac points can merge at the points other than
$\mathbf{M}_1$, $\mathbf{M}_2$, $\mathbf{M}_3$ and $\boldsymbol{\Gamma}$. 
The topological phase transitions caused by these merging Dirac points
are shown by the orange lines in Fig.~\ref{figparam4x},
which we will 
discuss in Section \ref{section6}.

The boundary lines 
$M_1$ and $M_3$ in $t_b/t_a - t_c/t_a$ plane
move closer and the regions $R_1$ and $R_2$ becomes
narrower, as $t_{3a}+t_{3b}+t_{3c}$  becomes large,
as seen in Fig.~\ref{figfig6}~(a).
 When $t_{3a}+t_{3b}+t_{3c}=t_a$, the boundaries ($M_1$ and $M_3$)
in the $t_b/t_a - t_c/t_a$ plane 
overlap and 
the regions $R_1$ and $R_3$ disappear in the phase diagram 
in the $t_b/t_a - t_c/t_a$ plane.
If $t_{3a}+t_{3b}+t_{3c}>t_a$, $M_1$ and $M_3$ lines are exchanged
and the new region $R_0$ appears,
as seen in Fig.~\ref{figfig6}~(b).

\section{Moving Dirac points with only nearest neighbor hoppings}
\label{section5}
In this section, only nearest neighbor hoppings are considered
($t^{\prime}_a=t^{\prime}_b=t^{\prime}_c=t^{\prime\prime}_a
=t^{\prime\prime}_b=t^{\prime\prime}_c=0$ 
and $t_{3a}=t_{3b}=t_{3c}=0$). 
When $|t_a|$, $|t_b|$ and $|t_c|$ satisfy the ``triangle inequality'',
\begin{equation}
 \left| \frac{|t_b|}{|t_a|} -1 \right| 
\leq \frac{|t_c|}{|t_a|} 
\leq \frac{|t_b|}{|t_a|} +1,
\label{eqtriangle}
\end{equation}
there are two Dirac points at $\mathbf{k}=\pm \mathbf{k}^{*}$.
The position of the Dirac points is obtained by 
the equations\cite{Hasegawa2006,Kishigi2008,Kishigi2008b,Kishigi2011}
\begin{align}
 \cos(\mathbf{v}_1 \cdot \mathbf{k}^*) 
&= \frac{t_c^2-t_a^2-t_b^2}{2 t_a t_b}, 
\label{eqeq3a}\\
 \cos(\mathbf{v}_2 \cdot \mathbf{k}^*) 
&= \frac{t_b^2-t_a^2-t_c^2}{2 t_c t_a}, 
\label{eqeq3b}\\
 \cos((\mathbf{v}_1-\mathbf{v}_2) \cdot \mathbf{k}^*) 
&= \frac{t_a^2-t_b^2-t_c^2}{2 t_b t_c}. 
\label{eqeq3c}
\end{align}
Note that the third equation (Eq. (\ref{eqeq3c})) is satisfied if 
the first and the second equations (Eqs. (\ref{eqeq3a}) and (\ref{eqeq3b}))
are satisfied and
\begin{equation}
t_b t_c \sin (\mathbf{v}_1 \cdot \mathbf{k}^*)
\sin (\mathbf{v}_2 \cdot \mathbf{k}^*) > 0 .
\end{equation}
The condition, Eq.~(\ref{eqtriangle}), is fulfilled 
in the regions $R_1$, $R_2$, $R_3$, $R_4$, $R_1^*$, $R_2^*$, 
$R_3^*$, and $R_4^*$,
in Fig.~\ref{figparam013d}~(a). 
%
\begin{figure}[bt]
\begin{center}
\vspace{0.8cm}
\includegraphics[width=0.35\textwidth]{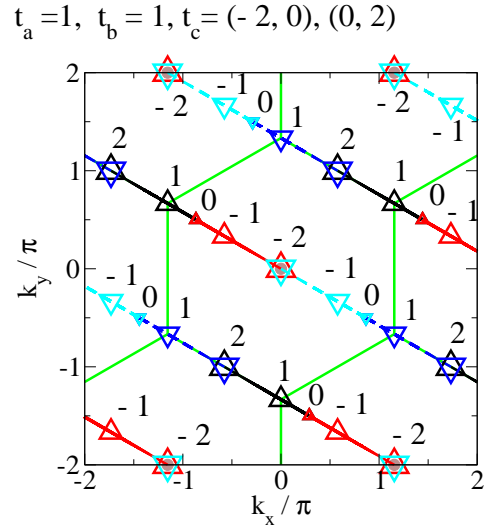}
\end{center}
\caption{(Color online)
Locations and trajectories of the Dirac points as a function of 
$t_c$ with $t_a=t_b=1$ being fixed. Only nearest neighbor hoppings
are considered. 
Solid and Broken lines are the trajectories of the Dirac points,
which are converted by the space inversion each other.
Up triangles and down triangles are the position of the 
Dirac points at $t_c$ indicated at the figure. When $t_c=2$
and $t_c=-2$, two Dirac points merge at the 
$M_1$ point $(\frac{\pi}{\sqrt{3}},\pi)$
and the $\Gamma$ point $(0,0)$, respectively.
When $t_c=0$ the system becomes one-dimensional. Therefore,
the smallest up and down triangles are not the Dirac points,
but Dirac points approach these points as $t_c \to 0$.}
\label{figfig1}
\end{figure}
\begin{figure}[bt]
\begin{center}
\vspace{1cm}
\includegraphics[width=0.35\textwidth]{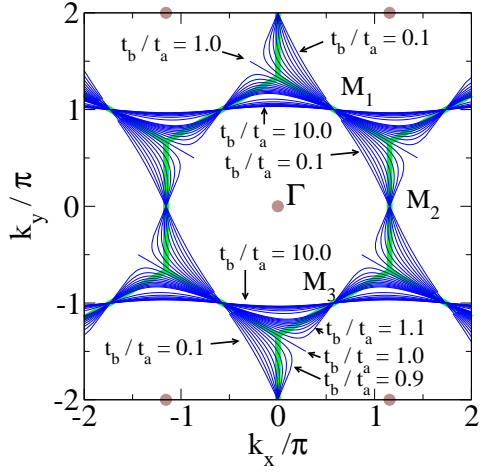}
\end{center}
\caption{(Color online)
Locations of the Dirac points in the anisotropic 
honeycomb lattice with $t_a=1$, $t_b=0.1, 0.2, \cdots 0.9, 1.0, 1.1, 
\cdots 1.9, 2.0, 3.0, 4.0, \cdots 9.0, 10.0$.
The lines are the trajectories of the Dirac points for $|t_b-t_a| < t_c
< t_b+t_a$ (region $R_1$ in Fig.~\ref{figparam013d})(a). }
\label{figfig00}
\end{figure}
\begin{figure}[bt]
\begin{center}
\vspace{1cm}
\includegraphics[width=0.40\textwidth]{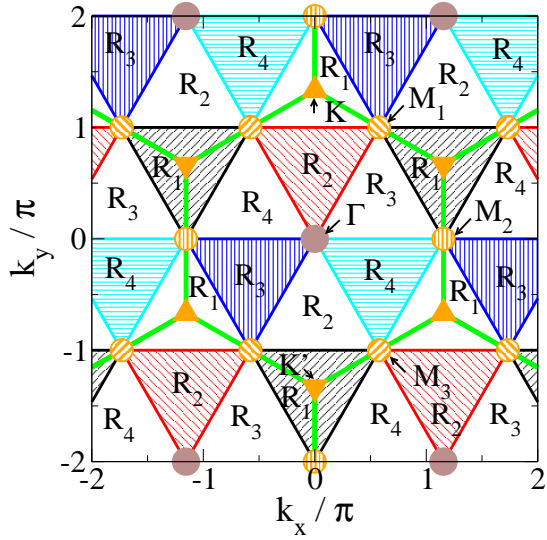}
\end{center}
\caption{(Color online)
Divided regions of the moving Dirac points 
in the case of $(t_a, t_b, t_c)$ 
being in the regions of $R_1$, $R_2$,
$R_3$ and $R_4$ in Fig.~\ref{figparam013d}(a).
}
\label{figfig0}
\end{figure}
We plot the trajectories of the Dirac points as $t_c$ is changed
for $t_a=t_b=1$ in Fig.~\ref{figfig1}.
When $t_c=1$ two Dirac points are located at $\mathbf{K}$ and $\mathbf{K}'$
(up triangles and down triangles labeled by ``1'' in Fig.~\ref{figfig1}).
The Dirac points move to $\mathbf{M}_1$ and the equivalent points
 as $t_c$ increases
(blue broken lines and black lines in Fig.~\ref{figfig1}),
 and they merge 
at $\mathbf{M}_1$ and the equivalent points when $t_c=2$. 
Dirac points annihilate when $t_c>2$.
Merging of two Dirac points causes a topological phase transition 
and this phase boundary is shown 
as $M_1$ in Fig.~\ref{figparam013d} and Fig.~\ref{figpara2d}.
On the other hand, when $t_c$ is decreased, Dirac points move in the opposite
direction. Dirac points approach to 
the midpoint of $\mathbf{M}_2$ and $\mathbf{M}_3$, 
$\pm (\mathbf{M}_2+\mathbf{M}_3)/2
=\pm (\sqrt{3}\pi/2,-\pi/2)$.
When $t_c=0$, the system becomes one-dimensional and two bands intersect
on the line.  
The smallest triangles with the label 
``0'' in Fig.~\ref{figfig1} should be considered as a singular
 points for the location of Dirac points as $t_c \to 0$.
As seen in Fig.~\ref{figparam013d}~(a) and Fig.~\ref{figpara2d}, 
$t_a=t_b=1$ and $t_c=0$
is the line (in Fig.~\ref{figparam013d}~(a))  
or the point (in Fig.~\ref{figpara2d}) of intersection 
for the boundary planes or lines
of $M_2$ and $M_3$. 
 When $t_c$ is changed to be negative, 
the parameters move into the region $R_4$ in Fig.~\ref{figparam013d}~(a)
and in Fig.~\ref{figpara2d}.
Dirac points move on the cyan broken lines and the red lines
in Fig.~\ref{figfig1} and merge at $\boldsymbol{\Gamma}$ 
when $t_c=-2$.

Similarly, if we change $t_a$ with fixing $t_b=t_c=1$,
Dirac points move on the vertical lines,
i.e., the figure is rotated by 120 degree.  
In that case Dirac points merge at $\mathbf{M}_2$ and 
the equivalent points when $t_a=2$.

In Fig.~\ref{figfig00} we plot a set of trajectories
as $t_c$ is changed ($|t_b - t_a| \leq t_c \leq t_b+t_a$) 
for several values of $t_b/t_a >0$.
Dirac points are located in the triangular regions around 
$\mathbf{K}$ and $\mathbf{K}'$,
when the parameters $t_a$, $t_b$, and $t_c$ are
in the $R_1$ region in Fig.~\ref{figparam013d}. 
Similarly, we obtain that
the Dirac points move in other triangular regions in Fig.~\ref{figfig0}
when $t_a$, $t_b$, and $t_c$ are in $R_2$, $R_3$ or $R_4$ regions
in Fig.~\ref{figparam013d}.
In Fig.~\ref{figfig0} we plot the regions for the 
Dirac points when parameters $t_a$, $t_b$ and $t_c$ are in $R_1$, $R_2$, 
$R_3$ and $R_4$ in Fig.~\ref{figparam013d}.

\section{Dirac points 
in the system with third-nearest-neighbor hoppings}
\label{section6}
%
%
\begin{figure}[bt]
\begin{center}
\includegraphics[width=0.29\textwidth]{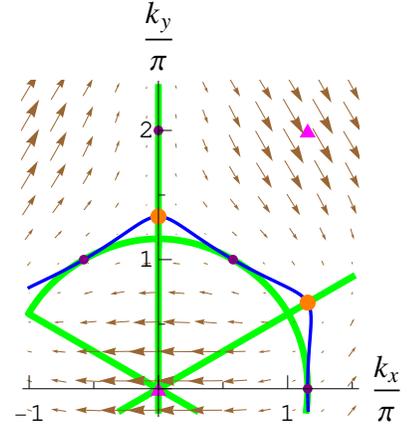}
\end{center}
\caption{(Color online)
Lines for $\epsilon_1(\mathbf{k})=0$ (thin blue lines)
and  $\epsilon_2(\mathbf{k})=0$ (thick green lines)
for $t_a=t_b=t_c=1$, $t_{3a}=t_{3b}=t_{3c}=1/3$.
Green and Blue lines touch at $\mathbf{M}_1$, $\mathbf{M}_2$ 
and $\mathbf{M}_3$.
}
\label{figmath033}
\end{figure}
\begin{figure}[bt]
\begin{center}
\includegraphics[width=0.29\textwidth]{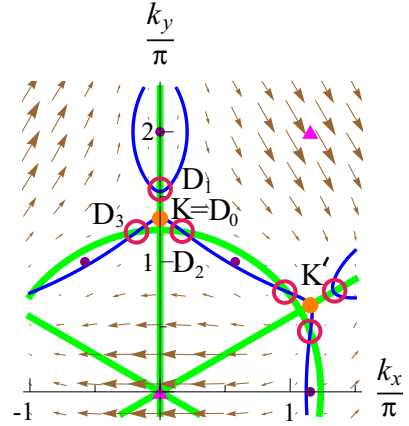}
\end{center}
\caption{(Color online)
Lines for $\epsilon_1(\mathbf{k})=0$ (thin blue lines)
and  $\epsilon_2(\mathbf{k})=0$ (thick green lines)
for $t_a=t_b=t_c=1$, $t_{3a}=t_{3b}=t_{3c}=0.4$.
Dirac points appear at $\mathbf{K}$ and $\mathbf{K}'$ and 
additional  points
(red circles, $\mathbf{D}_1$, $\mathbf{D}_2$ and $\mathbf{D}_3$).
The topological number is $-1$ and $+1$ for 
the additional three points around $\mathbf{K}$
 and $\mathbf{K}'$, respectively.
}
\label{figmath04}
\end{figure}
\begin{figure}[bt]
\begin{center}
\includegraphics[width=0.29\textwidth]{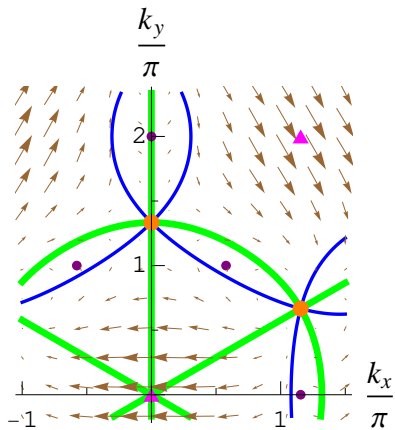}
\end{center}
\caption{(Color online)
Lines for $\epsilon_1(\mathbf{k})=0$ (thin blue lines)
and  $\epsilon_2(\mathbf{k})=0$ (thick green lines)
for $t_a=t_b=t_c=1$, $t_{3a}=t_{3b}=t_{3c}=0.5$.
Four Dirac points around $\mathbf{K}$ and $\mathbf{K}'$  merge 
and the topological number becomes $\mp 2$ at $\mathbf{K}$ 
and $\mathbf{K}'$, respectively.
}
\label{figmath05}
\end{figure}
\begin{figure}[bt]
\begin{center}
\includegraphics[width=0.29\textwidth]{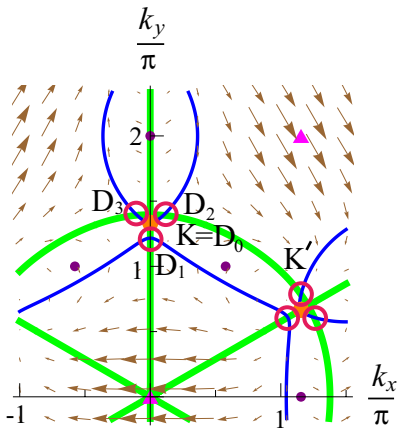}
\end{center}
\caption{(Color online)
Lines for $\epsilon_1(\mathbf{k})=0$ (thin blue lines)
and  $\epsilon_2(\mathbf{k})=0$ (thick green lines)
for $t_a=t_b=t_c=1$, $t_{3a}=t_{3b}=t_{3c}=0.6$.
Dirac points locate at the intersection points of green lines and blue lines,
i.e., $\mathbf{K}$, $\mathbf{K}'$, $\mathbf{D}_1$, 
$\mathbf{D}_2$ and $\mathbf{D}_3$. 
}
\label{figmath06}
\end{figure}
When the third-nearest-neighbor hoppings 
are finite, more than two Dirac points
are possible to exist.
For simplicity we assume all hoppings, 
$t_a$, $t_b$, $t_c$, $t_{3a}$, $t_{3b}$, and $t_{3c}$ are positive.
Generalization to the negative hoppings is straightforward.
\subsection{Symmetric case ($C_{6v}$)}
\label{sec6a}
In this subsection we discuss the symmetric system,
\begin{align}
t   &\equiv t_a = t_b = t_c, \\
t_3 &\equiv t_{3a} = t_{3b} = t_{3c}.
\end{align}
Bena and Simon\cite{Bena2011}  have  
studied this model. They have shown that new Dirac points appear
 when $t_3 >t/3$. 
As shown in  Fig.~\ref{figmath1},
the blue lines, which is  given by $\epsilon_1(\mathbf{k})=0$, and  
the green lines, which is  given by $\epsilon_2(\mathbf{k})=0$, 
have intersection points only at $\mathbf{K}$ and $\mathbf{K}'$,
when $t_3 < t/3$. Derivation is given in Appendix~\ref{refappA}.
The blue lines and the green lines
touch at $\mathbf{M}_1$, $\mathbf{M}_2$ and $\mathbf{M}_3$, 
when $t_3 = t/3$ (Fig.~\ref{figmath033}).
We perform power series expansions of $\epsilon_1(\mathbf{k})$ and  
$\epsilon_2(\mathbf{k})$ about $\mathbf{M}_2$  and obtain
\begin{align}
  \epsilon_1 (\mathbf{k}) &\approx 
 + t \left(  k_x-\frac{2\sqrt{3}\pi}{3} 
            - \frac{1}{24}   k_y^2 \right)  ,\\
  \epsilon_2(\mathbf{k}) &\approx
   \frac{\sqrt{3}}{3}t \left(  k_x-\frac{2\sqrt{3}\pi}{3} 
+\frac{1}{8}  k_y^2 \right),
\end{align}
when $t_3=t/3$.
This means that 
merged Dirac point exists at $\mathbf{M}_2$ when $t_3 = t/3$
and the energy is given by
\begin{equation}
\sqrt{(\epsilon_1 (\mathbf{k}))^2+(\epsilon_2 (\mathbf{k}))^2}
\approx
t \sqrt{ \frac{4}{3} \left(  k_x-\frac{2\sqrt{3}\pi}{3} \right)^2
 +\left( \frac{k_y^2}{12} \right)^2},
\end{equation}
i.e., the energy is proportional to $|k_x-k_x^*|$ in one direction and
proportional to $(k_y - k_y^*)^2$ in other direction.
The merged Dirac points, which have these energy dispersion,
are called semi-Dirac points\cite{Banerjee2009}.
When the semi-Dirac points exit, 
density of states is proportional to the square root of 
the energy\cite{Hasegawa2006}
\begin{equation}
 D(\epsilon) \propto \sqrt{\epsilon},
\end{equation}
and the energy of the Landau levels 
varies as $B^{2/3}$[Ref.~\onlinecite{Dietl2008}]. 

When $t_3$ is larger than $t/3$, 
each touching point of blue and green lines changes to a pair of 
the intersection points, i.e.,
each semi-Dirac point separate into 
two Dirac points with topological number $\pm 1$, and
there exist eight Dirac points, as shown
by $\mathbf{D}_1$, $\mathbf{D}_2$ and $\mathbf{D}_3$ 
around $\mathbf{K}$  ($= \mathbf{D}_0$) in Fig.~\ref{figmath04}.
As $t_3/t$ becomes large, two Dirac points stay at $\mathbf{K}$ 
and $\mathbf{K}'$ and 
other Dirac points move from $\mathbf{M}_1$, $\mathbf{M}_2$ and 
$\mathbf{M}_3$ to $\mathbf{K}$
and $\mathbf{K}'$.
Three Dirac points moving to $\mathbf{K}$ for $1/3 < t_3/t < 1/2$ 
($\mathbf{D}_1$, $\mathbf{D}_2$ and $\mathbf{D}_3$ 
in Fig.~\ref{figmath04}) have the topological number
$-1$ and three Dirac points moving to $\mathbf{K}'$ 
have topological number $+1$.
When $t_3=t/2$, four Dirac points merge at $\mathbf{K}$ and $\mathbf{K}'$.
In that case two blue lines and two green lines intersect at $\mathbf{K}$ 
and $\mathbf{K}'$,
as shown in Fig.~\ref{figmath05}.
The merged Dirac points at $\mathbf{K}$ and $\mathbf{K}'$ 
have topological number $-2$ 
and $+2$, respectively ($1 \times (+1) + 3 \times (-1) = -2$ at 
$\mathbf{K}$, for example).
The topological number at $\mathbf{K}$ can be also obtained as follows.
For $(k_x,k_y) \approx \mathbf{K} = (0, 4\pi/3)$,
we write 
\begin{align}
 k_x &= \kappa \cos (\theta), \\
 k_y-\frac{4}{3}\pi &= \kappa \sin (\theta).
\end{align}
When $t_3=t/2$, we obtain
\begin{align}
\epsilon_1(\mathbf{k}) & \approx   \frac{3}{8} t \kappa^2 \cos (2\theta) \\
\epsilon_2(\mathbf{k}) & \approx - \frac{3}{8} t \kappa^2 \sin (2\theta) .
\end{align}
Comparing these equations with Eqs.~(\ref{eqdefphi1}) and (\ref{eqdefphi2}),
we obtain 
\begin{equation}
 \phi(\mathbf{k}) = -2 \theta,
\end{equation}
and the topological number of the vector 
$(\epsilon_1(\mathbf{k}),\epsilon_1(\mathbf{k}))$
at $\mathbf{K}$ is $-2$.
The energy at this point is obtained as
\begin{equation}
 \sqrt{(\epsilon_1(\mathbf{k}))^2+(\epsilon_2(\mathbf{k}))^2}
\approx \frac{3}{8} t \left(
k_x^2 + \left(k_y-\frac{4}{3}\pi \right)^2 \right),
\end{equation}
which has been obtained by Bena and Simon\cite{Bena2011}.
In this case the density of states is constant  
near $\epsilon \approx 0$.

When $t_3>t/2$,  
eight Dirac points appear again. 
Two Dirac points are at $\mathbf{K}$ 
and $\mathbf{K}'$ having topological number $+1$ and $-1$,
respectively.
There exist three Dirac points with topological number $-1$ 
around $\mathbf{K}$ and three Dirac points with topological number $+1$ 
around $\mathbf{K}'$. They form the equilateral triangles 
as in the case of $t/3< t_3 < t/2$, but the triangles  
are upside-down, as shown in Fig.~\ref{figmath06} for $t_{3}=0.6 t$.
\subsection{direction-dependent hoppings with third-nearest neighbor hoppings}
 %
\begin{figure}[bt]
\begin{flushleft}\hspace*{0.5cm} (a)\end{flushleft}
\includegraphics[width=0.32\textwidth]{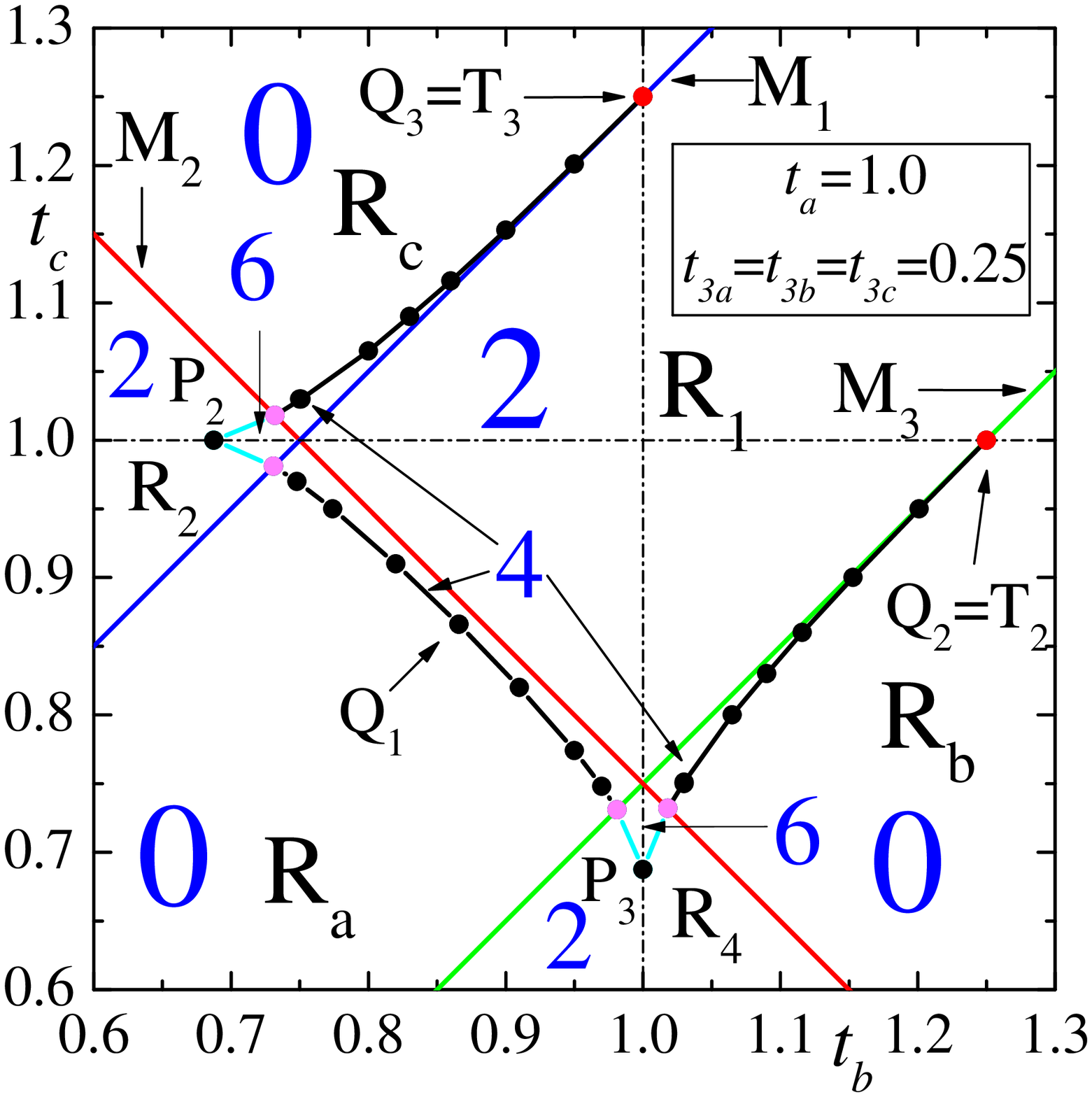}
\begin{flushleft}\hspace*{0.5cm} (b)\end{flushleft}
\includegraphics[width=0.32\textwidth]{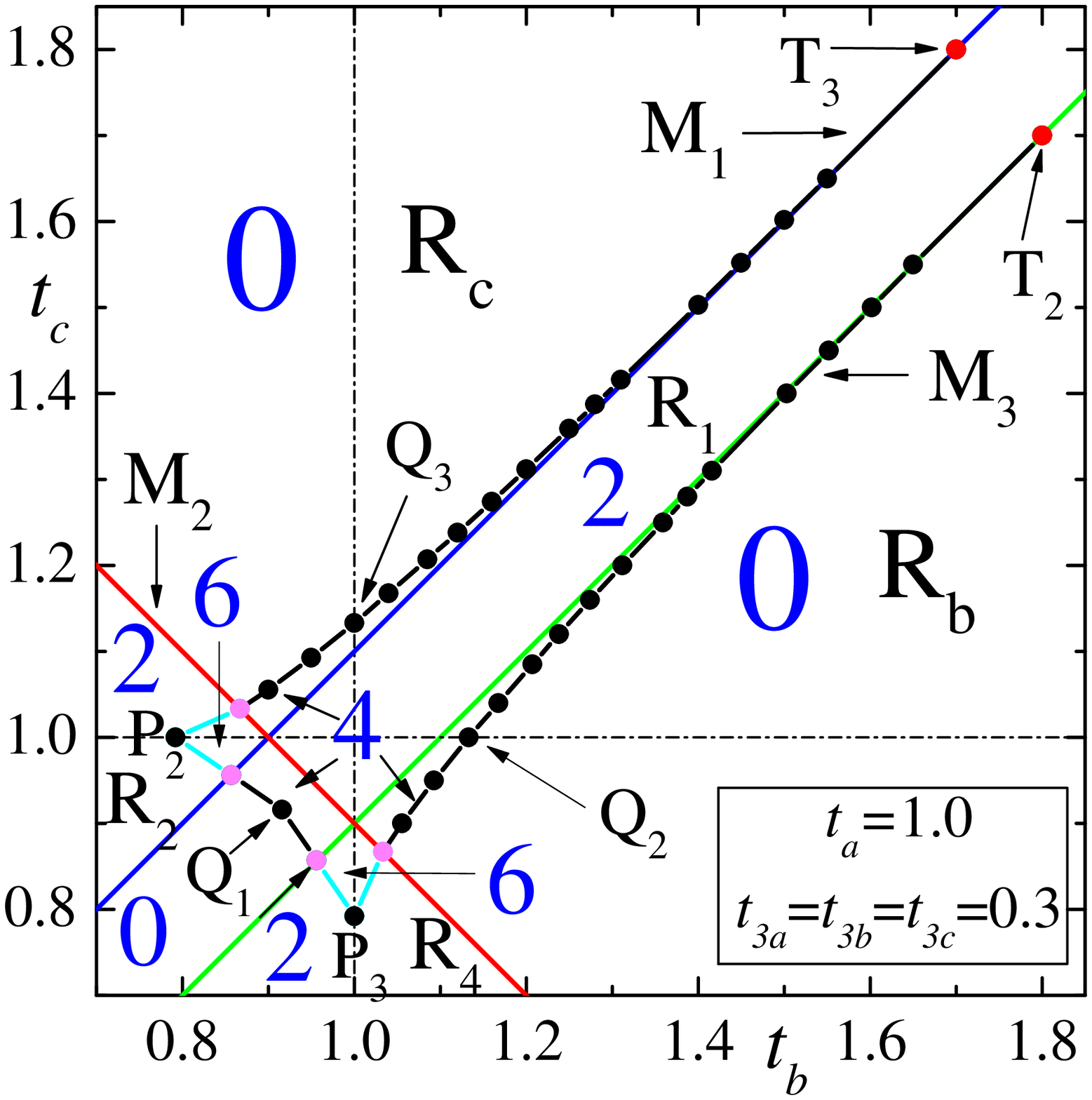}
\begin{flushleft}\hspace*{0.5cm} (c)\end{flushleft}
\includegraphics[width=0.32\textwidth]{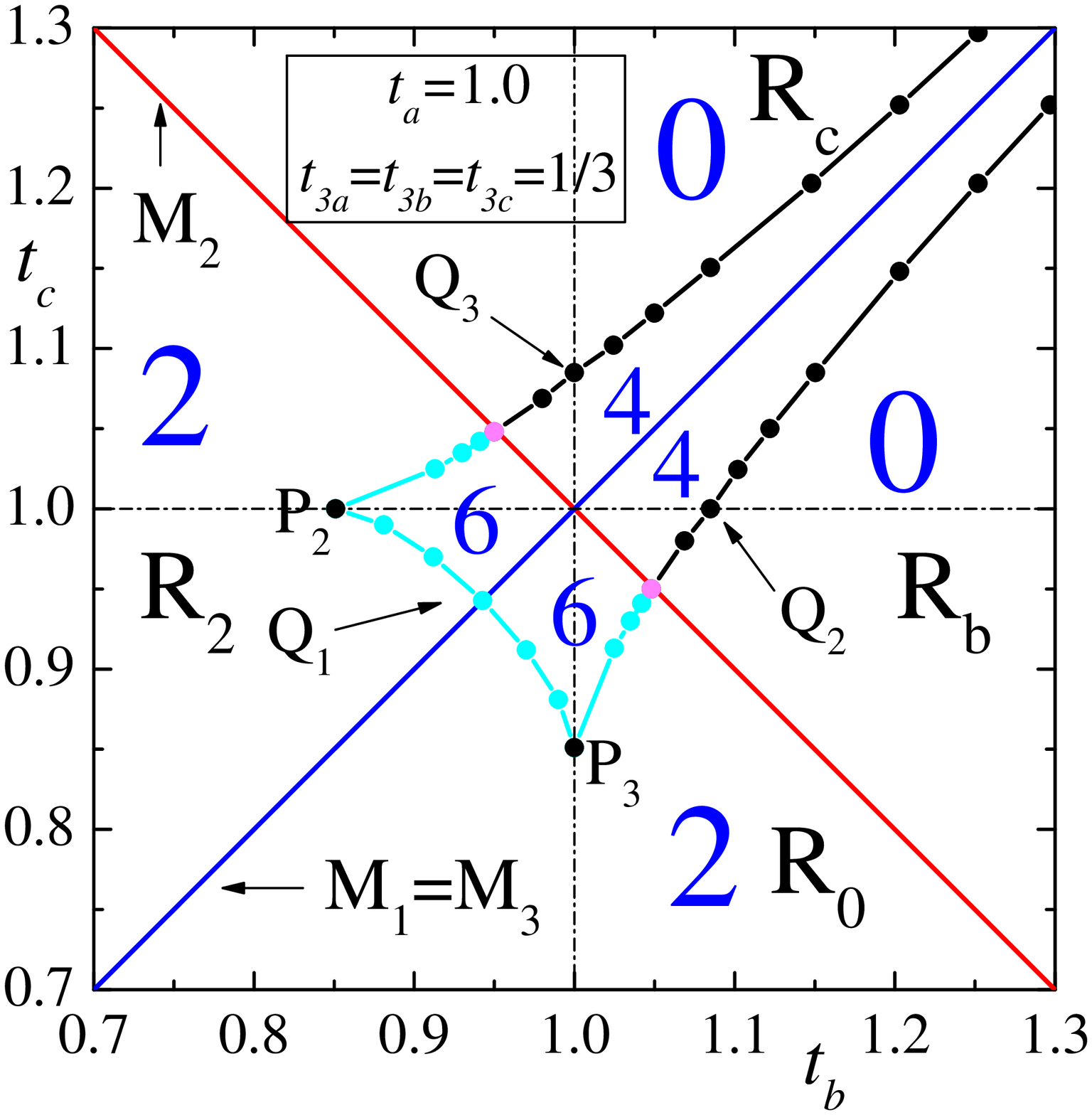}
\caption{(Color online)
Phase diagram in the $t_b$-$t_c$ plane 
obtained numerically at 
$t_a=1$ and $t_{3a}=t_{3b}=t_{3c}= 0.25$, $0.3$, and $1/3$. 
The numbers of the Dirac points in the Brillouin zone is
 shown in the figure.
}
\label{figphasenum}
\end{figure}
\begin{figure}[bt]
\begin{flushleft}\hspace*{0.5cm} (a)\end{flushleft}
\includegraphics[width=0.32\textwidth]{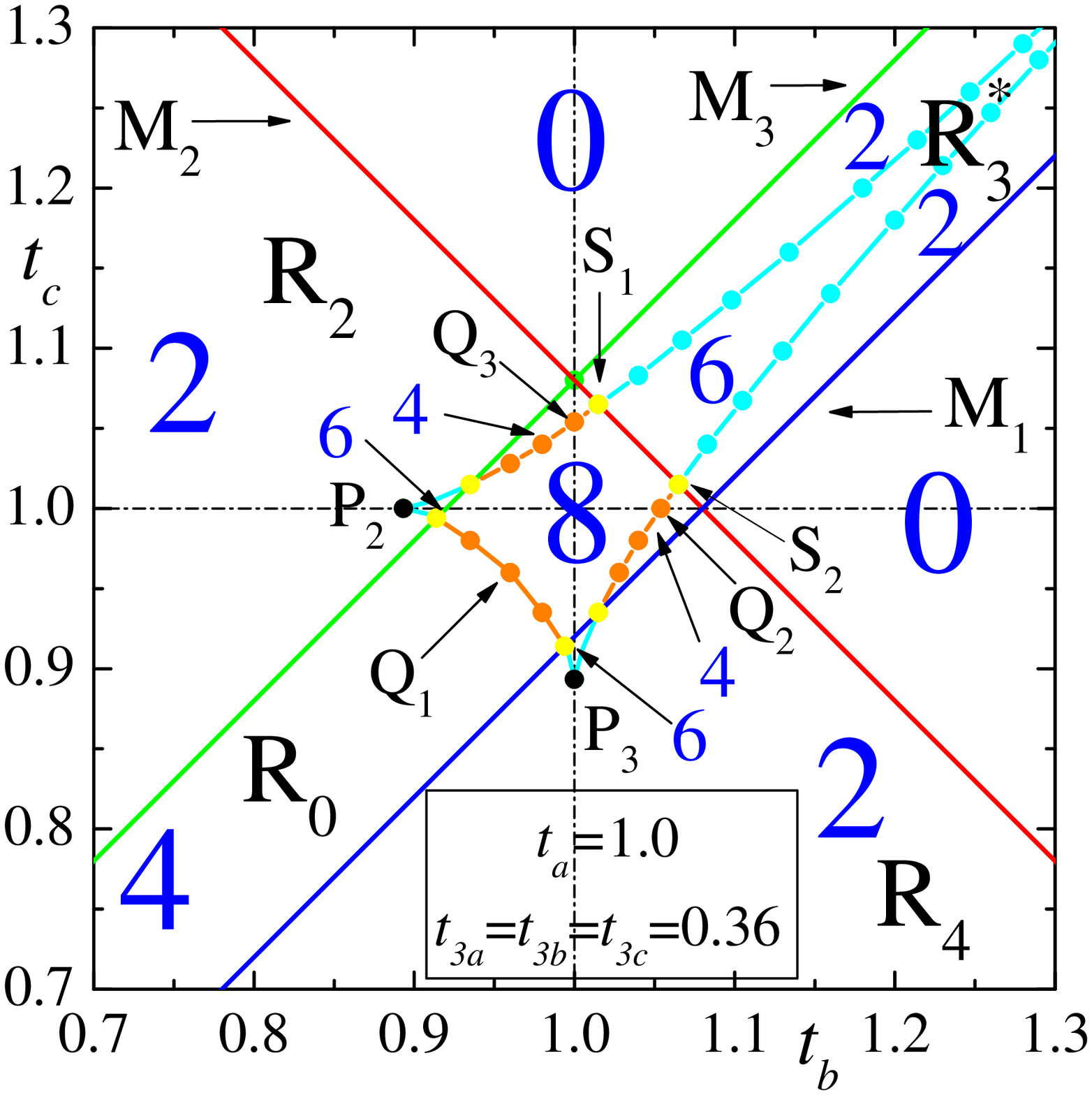}\\
\begin{flushleft}\hspace*{0.5cm} (b)\end{flushleft}
\includegraphics[width=0.32\textwidth]{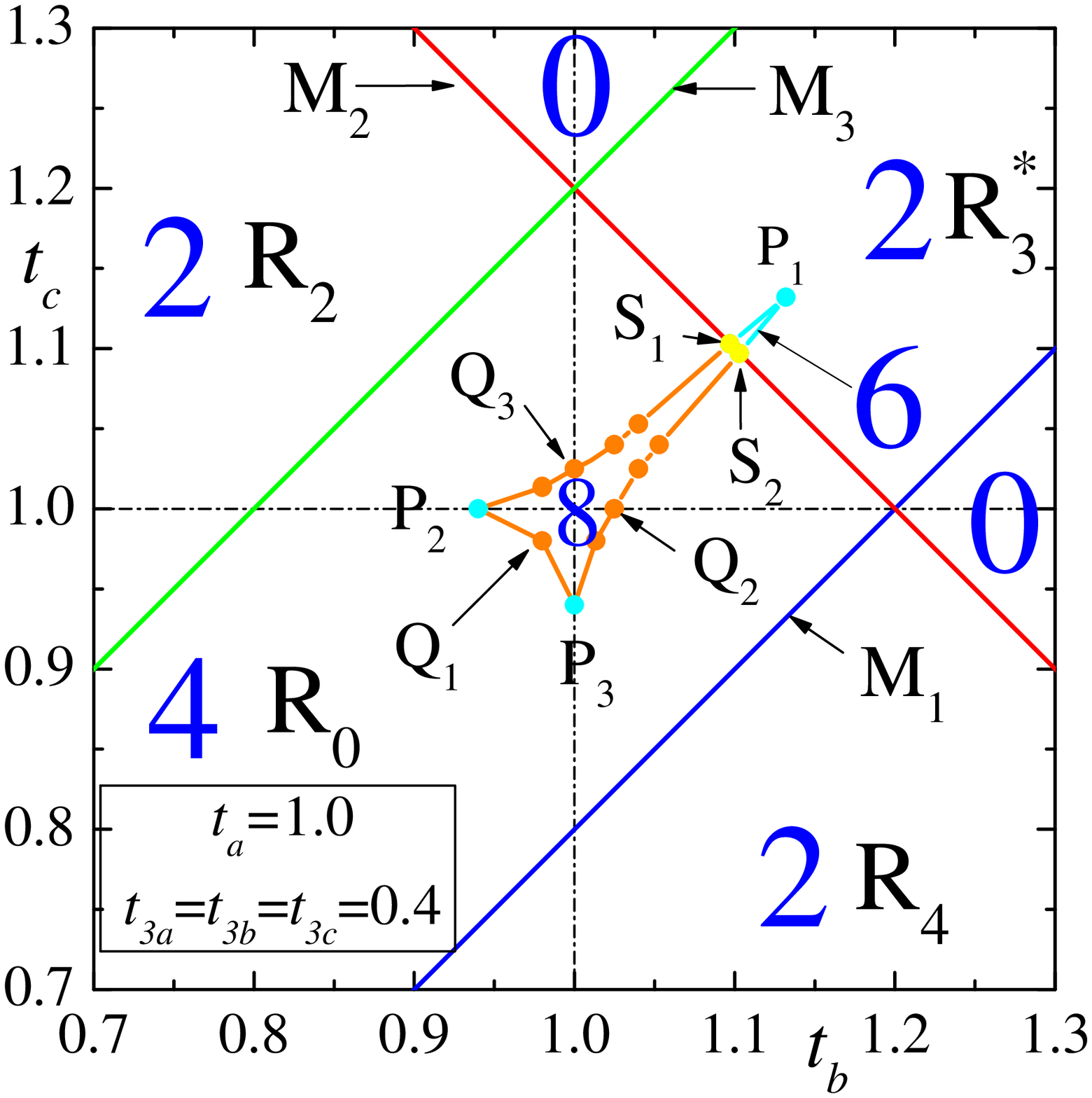}\\
\begin{flushleft}\hspace*{0.5cm} (c)\end{flushleft}
\includegraphics[width=0.32\textwidth]{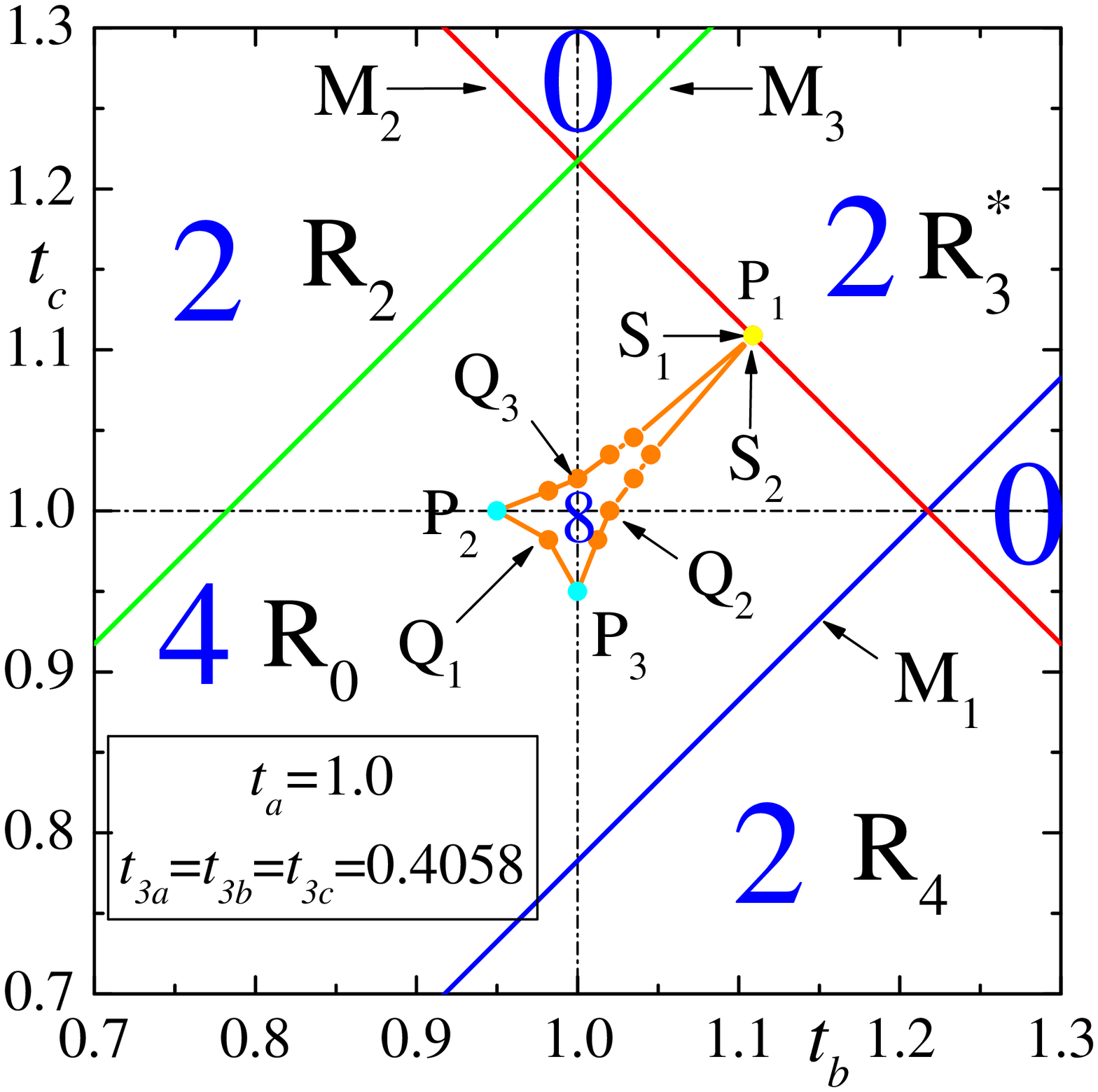}
\caption{(Color online)
Phase diagram in the $t_b$-$t_c$ plane at 
$t_a=1$ and $t_{3a}=t_{3b}=t_{3c}=0.36$, $0.4$ and $0.4058$.
 The numbers in the figure indicate the number of the Dirac points
in the Brillouin zone. 
}
\label{fig16_gx}
\end{figure}
If the hoppings depend on the direction, the phase diagram becomes
much richer. 
Even if  the hoppings between the third-nearest-neighbor sites
are small, there appear new phases 
as shown in Fig.~\ref{figparam4x}.
Two Dirac points can merge  at the point other than 
$\mathbf{M}_1$, $\mathbf{M}_2$, $\mathbf{M}_3$, and $\boldsymbol{\Gamma$}.
When a pair of the Dirac points merge at $\mathbf{k}^*$,
which are neither  
$\mathbf{M}_1$, $\mathbf{M}_2$, $\mathbf{M}_3$ nor $\boldsymbol{\Gamma$},  
in the Brillouin zone, other pair should
merge at  $-\mathbf{k}^*$ simultaneously,
 since we study the system with 
time-reversal symmetry.
Therefore, when  the parameters are moved across these phase boundary lines
(orange lines in Fig.~\ref{figparam4x} and curved lines in Fig.~\ref{figphasenum} 
and Fig.~\ref{fig16_gx}),
the number of the Dirac points changes by four, i.e.,
 $0 \leftrightarrow 4$, 
$2 \leftrightarrow 6$ and $4 \leftrightarrow 8$.

When the system has the reflection symmetry ($C_{2v}$ symmetry)
we can obtain the analytical 
expressions for some special points 
($\mathbf{P}_1$, $\mathbf{P}_2$, $\mathbf{P}_3$,
$\mathbf{Q}_1$, $\mathbf{Q}_2$, and $\mathbf{Q}_3$) in the phase diagram
(see Appendix~\ref{refappB}).
At $\mathbf{P}_1$, $\mathbf{P}_2$ and $\mathbf{P}_3$ in the phase diagram in 
$t_b/t_a - t_c/t_a$ plane,
three Dirac points merge. 

There exist
the tricritical points, at which  three phases with zero, two and
four Dirac points terminate,  in the parameter space
in $t_b$ and $t_c$, which we label  as $\mathbf{T}_1$, $\mathbf{T}_1'$, 
$\mathbf{T}_2$ and $\mathbf{T}_3$.

In order to make the discussion simpler, 
we assume the third-nearest-neighbor hoppings are independent of the direction,
i.e., we take $t_3 \equiv t_{3a}=t_{3b}=t_{3c}$. 
We plot $\mathbf{P}_1$, $\mathbf{P}_2$, $\mathbf{P}_3$, 
$\mathbf{T}_1$, $\mathbf{T}_1'$, $\mathbf{T}_2$
and $\mathbf{T}_3$ as lines in the 3D plot in $t_a -t_b - t_c$ space
in Fig.~\ref{figfig6}.  
In the $t_b-t_c$ plane at $t_a=1$,
these points are obtained as
\begin{align}
 \mathbf{P}_1 &= \left( \frac{2 t_3^{3/2}}{\sqrt{3 t_3-1}},
                        \frac{2 t_3^{3/2}}{\sqrt{3 t_3-1}} \right), 
  \label{eqeqp10} \\
 \mathbf{P}_2 &= \left( 3 t_3 - 4 t_3^3, 1 \right) ,
\label{eqeqp20}
\end{align}
and
\begin{equation}
 \mathbf{P}_3 = \left( 1, 3 t_3 - 4 t_3^3 \right) .
\label{eqeqp30}
\end{equation}
Note that while 
$\mathbf{P}_1$ exists only for $t_3 > 1/3$, $\mathbf{P}_2$ and 
$\mathbf{P}_3$ exist for any value of $t_3 \ne 0$.
The points $\mathbf{Q}_1$, $\mathbf{Q}_2$ and $\mathbf{Q}_3$ 
are given in Appendix \ref{appendixB2}.
We find that $\mathbf{Q}_1$ exists when $t_3>1/5$,
 and $\mathbf{Q}_2$ and $\mathbf{Q}_3$ exist when $t_3>1/4$.

 The tricritical points are given as
\begin{align}
 \mathbf{T}_1  
=& \Biggl(
 \frac{1+3 t_3}{2} + \frac{1+t_3}{2} \sqrt{\frac{1-5 t_3}{1+3 t_3}},
\nonumber \\
& \frac{1+3 t_3}{2} - \frac{1+t_3}{2} \sqrt{\frac{1-5 t_3}{1+3 t_3}}
 \Biggr) ,
\label{eqT1}
\\
 \mathbf{T}_1'&=\left( (\mathbf{T}_1)_y,(\mathbf{T}_1)_x \right),
\label{eqT1p}
\\
  \mathbf{T}_2 
=& \Biggl(
 \frac{ 1 - 3 t_3}{2} + \frac{1-t_3}{2} \sqrt{\frac{1+5 t_3}{1-3 t_3}},
\nonumber \\
& \frac{-1 + 3 t_3}{2} + \frac{1-t_3}{2} \sqrt{\frac{1+5 t_3}{1-3 t_3}}
 \Biggr) ,
\label{eqT2}
\end{align}
and
\begin{align}
 \mathbf{T}_3&=\left( (\mathbf{T}_2)_y,(\mathbf{T}_2)_x \right).
\label{eqT3}
\end{align}
Note that
$\mathbf{T}_1$ and $\mathbf{T}_1'$ exist when $0 < t_3 < 1/5$,
and $\mathbf{T}_2$ and $\mathbf{T}_3$ exist when $0 < t_3 < 1/3$.

The phase transition lines connections $\mathbf{P}_i$, $\mathbf{Q}_i$ 
and $\mathbf{T}_i$
are obtained numerically and  shown in Figs.~\ref{figphasenum} 
and \ref{fig16_gx}
for $t_3=0.25$, $0.3$, $1/3$, $0.36$, $0.4$ and $0.4058$.
The numbers of the Dirac points in the Brillouin zone changes by 2 when 
the phases are divided by the straight line and it changes by 4
when the phases are separated by the curved line.

\begin{figure}[bt]
\begin{center}
\includegraphics[width=0.48\textwidth]{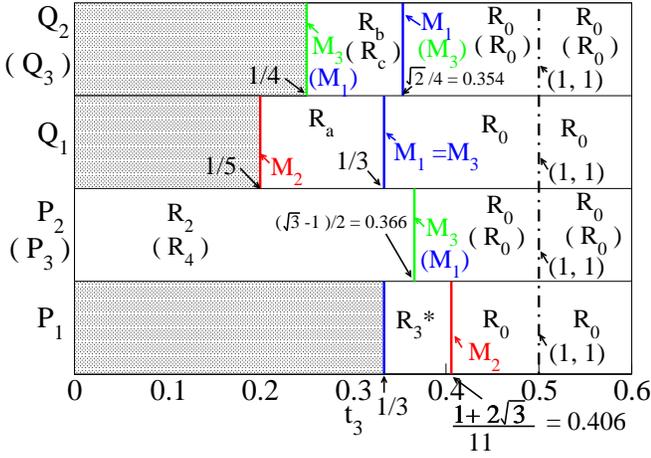}
\end{center}
\caption{(Color online)
Locations of the parameters for three Dirac points to merge
($\mathbf{P}_1$, $\mathbf{P}_2$ and $\mathbf{P}_3$) 
and for two Dirac points to merge
in the axisymmetric case
($\mathbf{Q}_1$, $\mathbf{Q}_2$ and $\mathbf{Q}_3$). 
When $t_3=0.5$ all points are at (1,1).
They do not exist 
in the shaded region
($0 < t_3 < 1/5$ for $\mathbf{Q}_1$, 
$0 < t_3 < 1/4$ for $\mathbf{Q}_2$ and $\mathbf{Q}_3$,
and $0 < t_3 < 1/3$ for $\mathbf{P}_1$).      
The regions labeled as  
$R_a$, $R_b$, $R_c$, $R_0$, $R_2$, $R_4$ and $R_3^*$,
are those shown in Fig.~\ref{figparam013d}. 
The vertical lines labeled as $M_1$, $M_2$ and $M_3$ show
that $\mathbf{P}_i$ and $\mathbf{Q}_i$ ($i=1$, $2$, and $3$) 
are on the
boundaries ($M_1$, $M_2$ and $M_3$, respectively).
}
\label{figparam00}
\end{figure}

\begin{figure}[bt]
\begin{center}
\includegraphics[width=0.48\textwidth]{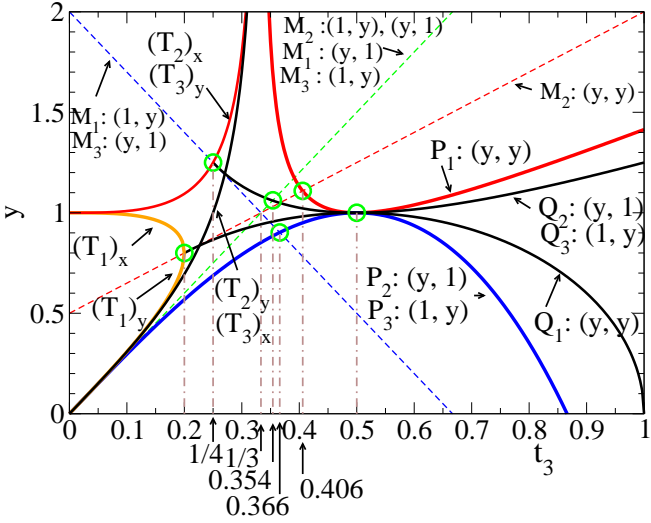}
\end{center}
\caption{(Color online)
Coordinates of the special points in the
phase diagram in $t_b/t_a$ - $t_c/t_a$
plane as a function of $t_3$ 
($t_3=t_{3a}=t_{3b}=t_{3c}$ and $t_a=1$).
At $\mathbf{P}_{1}$, $\mathbf{P}_2$ and $\mathbf{P}_3$
(Eqs.~(\ref{eqeqp10}), (\ref{eqeqp20}), and (\ref{eqeqp30})),
 three Dirac points merge, and 
at $\mathbf{Q}_1$, $\mathbf{Q}_2$ and $\mathbf{Q}_3$ 
(Eqs.~(\ref{eqeqq1}), (\ref{eqeqq2}), and (\ref{eqeqq3})),
two Dirac points merge
on the symmetric line ($k_x=0$) in the Brillouin zone.
}
\label{figcritical}
\end{figure}

In Fig.~\ref{figparam00} we show the regions where  $\mathbf{P}_i$ and
 $\mathbf{Q}_i$ locate. 
We plot the $t_3$ dependences of the 
coordinates of $\mathbf{P}_i$, $\mathbf{Q}_i$ 
and $\mathbf{T}_i$ in Fig.~\ref{figcritical}.

At the tricritical points, four Dirac points with topological numbers,
$+1$, $+1$, $-1$ and $-1$, 
merge at one of $\mathbf{M}_i$ in the Brillouin zone.
 
The density of states are calculated in Appendix \ref{appendixD}.
We obtain
\begin{equation}
 D(\epsilon) \propto \left\{
\begin{array}{ll}
|\epsilon|      & \mbox{at Dirac point}\\
\sqrt{|\epsilon|} & \mbox{when two Dirac points merge}\\
                & \mbox{ (semi-Dirac point) }\\
|\epsilon|^{\frac{1}{3}} &  \mbox{when three Dirac points merge}\\
                         &  \mathbf{P}_1, \mathbf{P}_2, \mathbf{P}_3\\
|\epsilon|^{\frac{1}{4}} &  \mbox{at tricritical point}\\
          &  \mathbf{T}_1, \mathbf{T}_1', \mathbf{T}_2, \mathbf{T}_3 .
\end{array} \right. 
\end{equation}
Note that
although the topological number of the
merged point of three Dirac points is the same as
that of the ordinary Dirac point, energy dependences
of the density of states are different.

%

\begin{figure}[bt]
\includegraphics[width=0.33\textwidth]{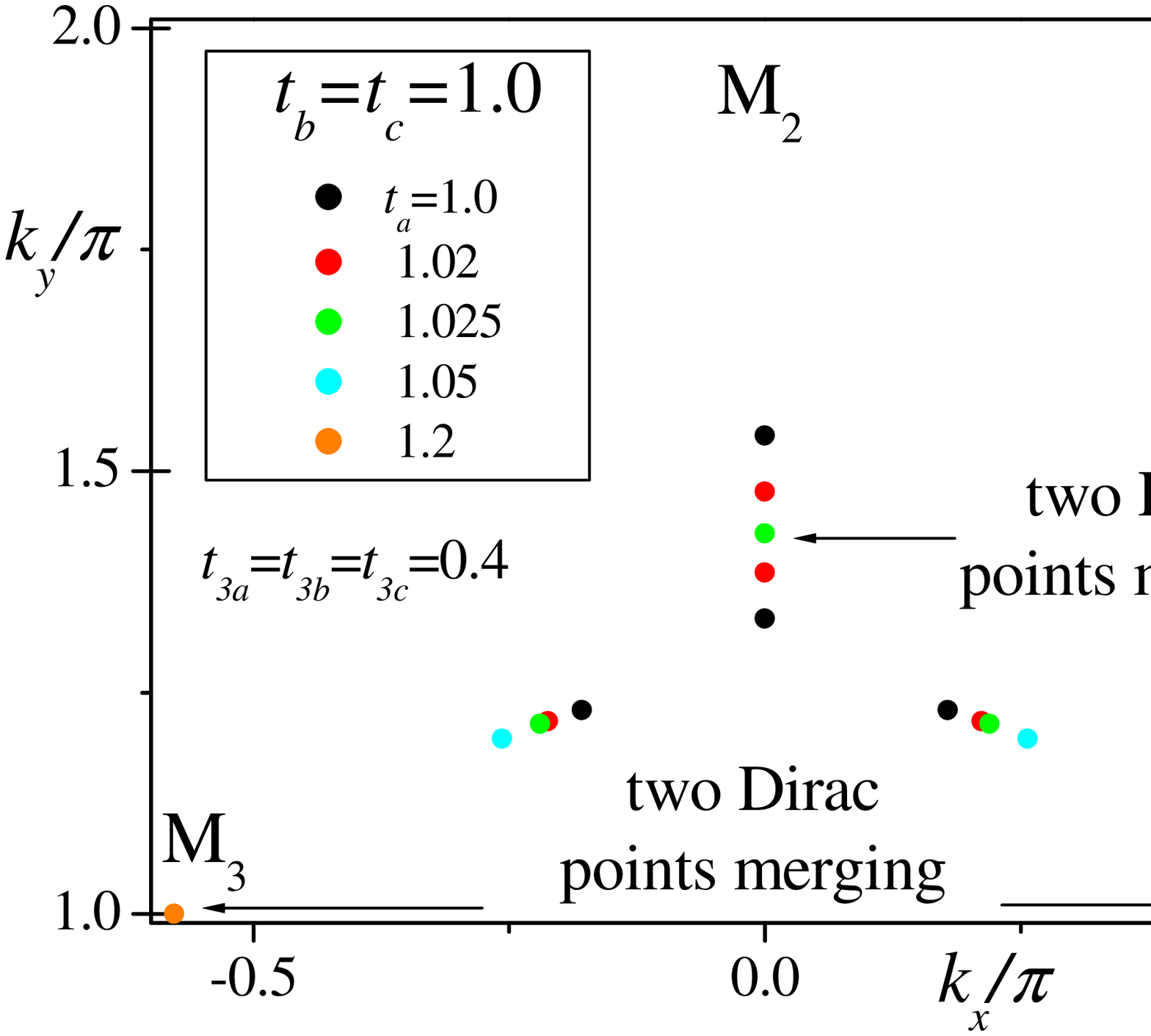}\\
\includegraphics[width=0.33\textwidth]{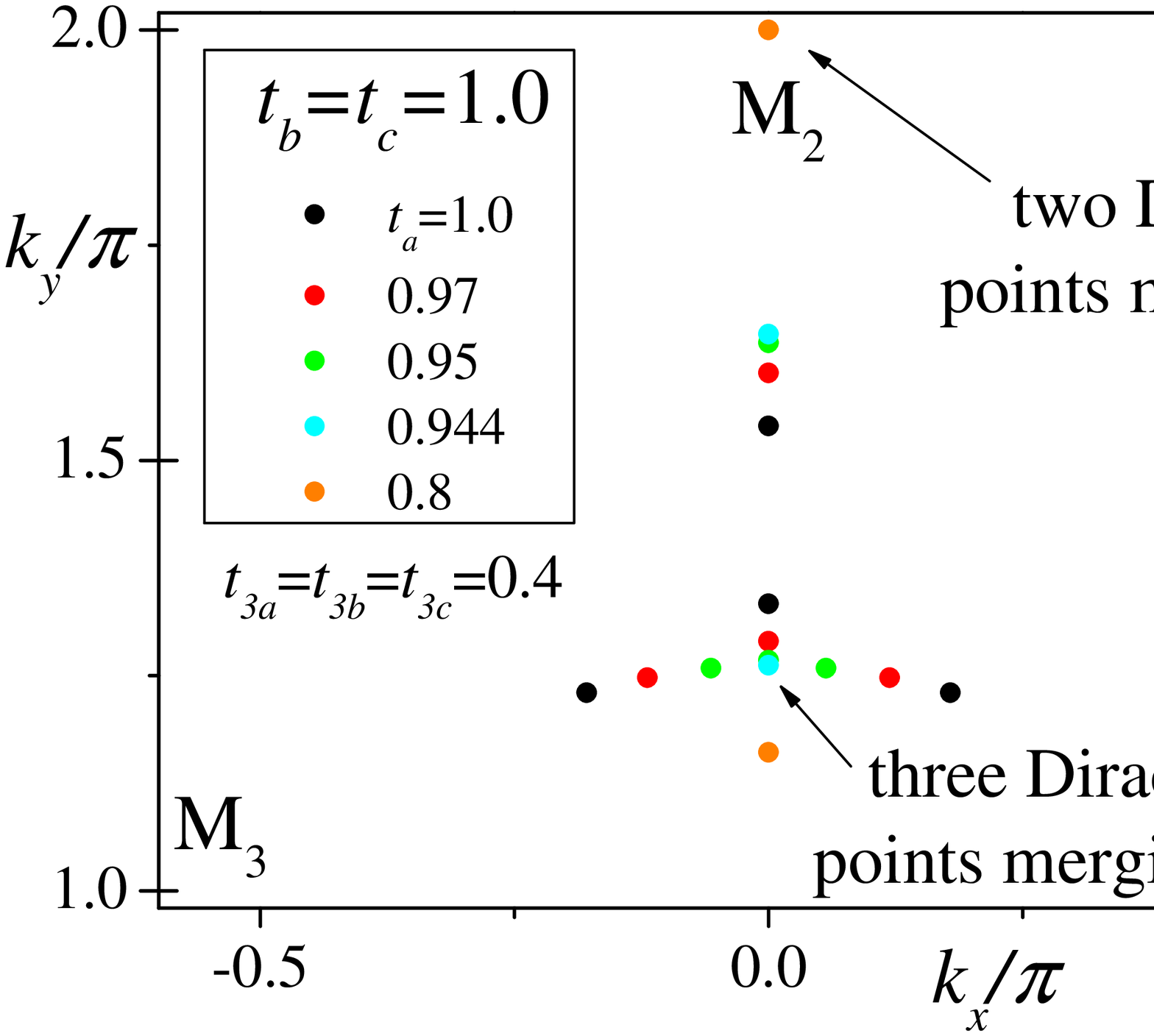}
\caption{(Color online)
The moving Dirac points by changing $t_a$
with fixed $t_b = t_c =1$ and $t_{3a}=t_{3b}=t_{3c}=0.4$. 
}
\label{fig16_e}
\end{figure}

\begin{figure}[bt]
\includegraphics[width=0.33\textwidth]{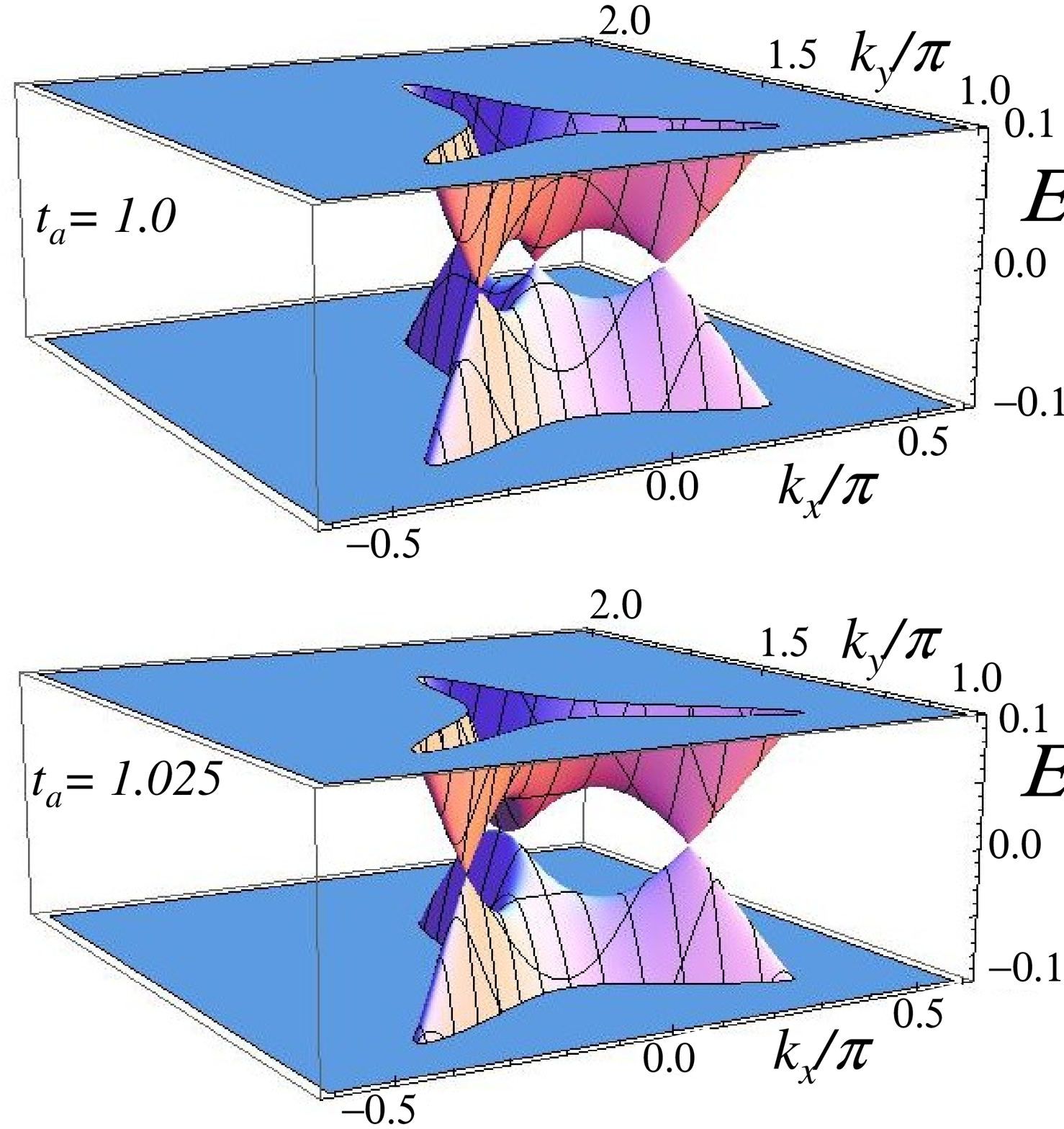}\\
\includegraphics[width=0.33\textwidth]{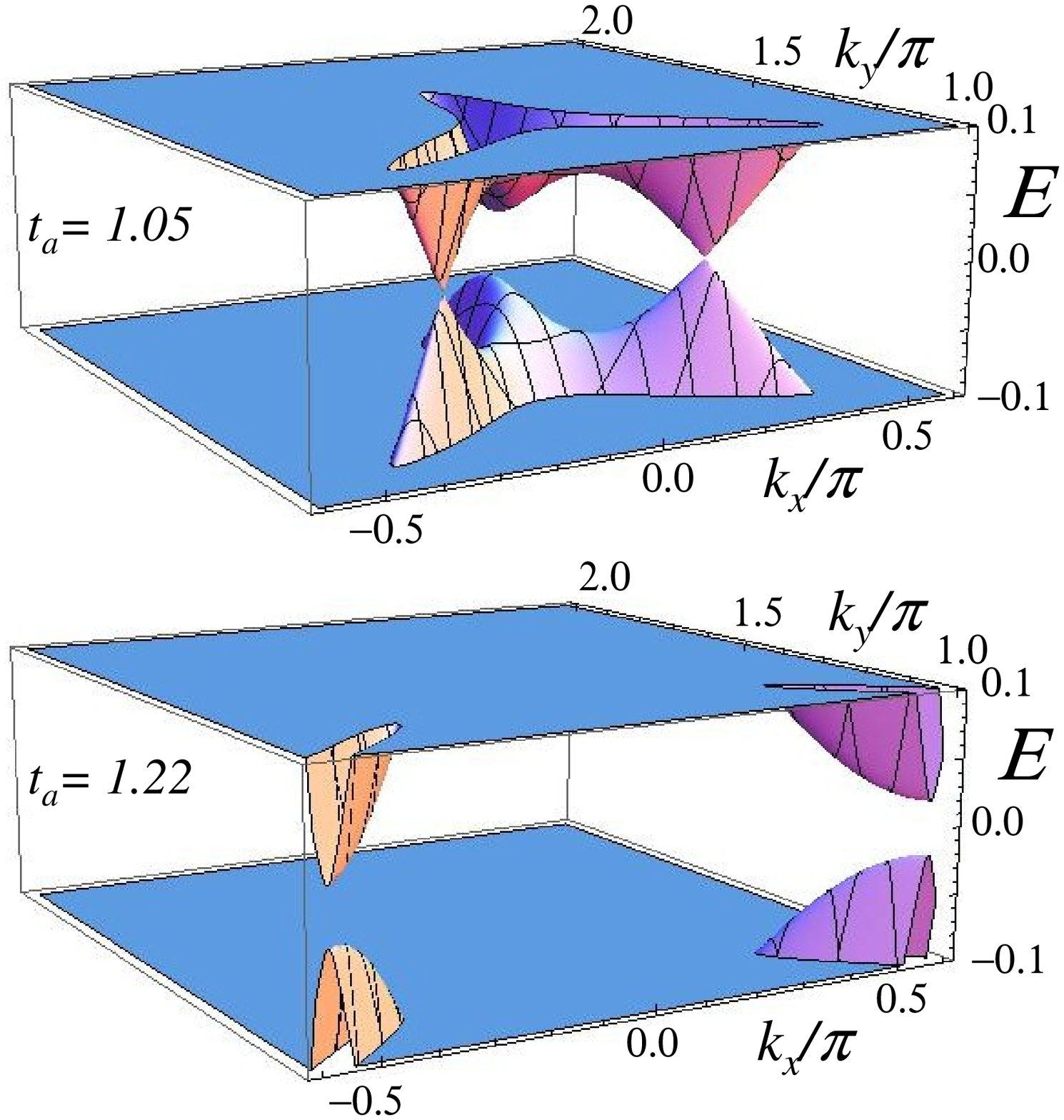}
\caption{(Color online)
3D plot of the energy as a function of the wave number
 for the parameters in Fig.\ref{fig16_e}
($t_a \geq 1$).
}\label{fig16_f}
\end{figure}

\begin{figure}[bt]
\includegraphics[width=0.33\textwidth]{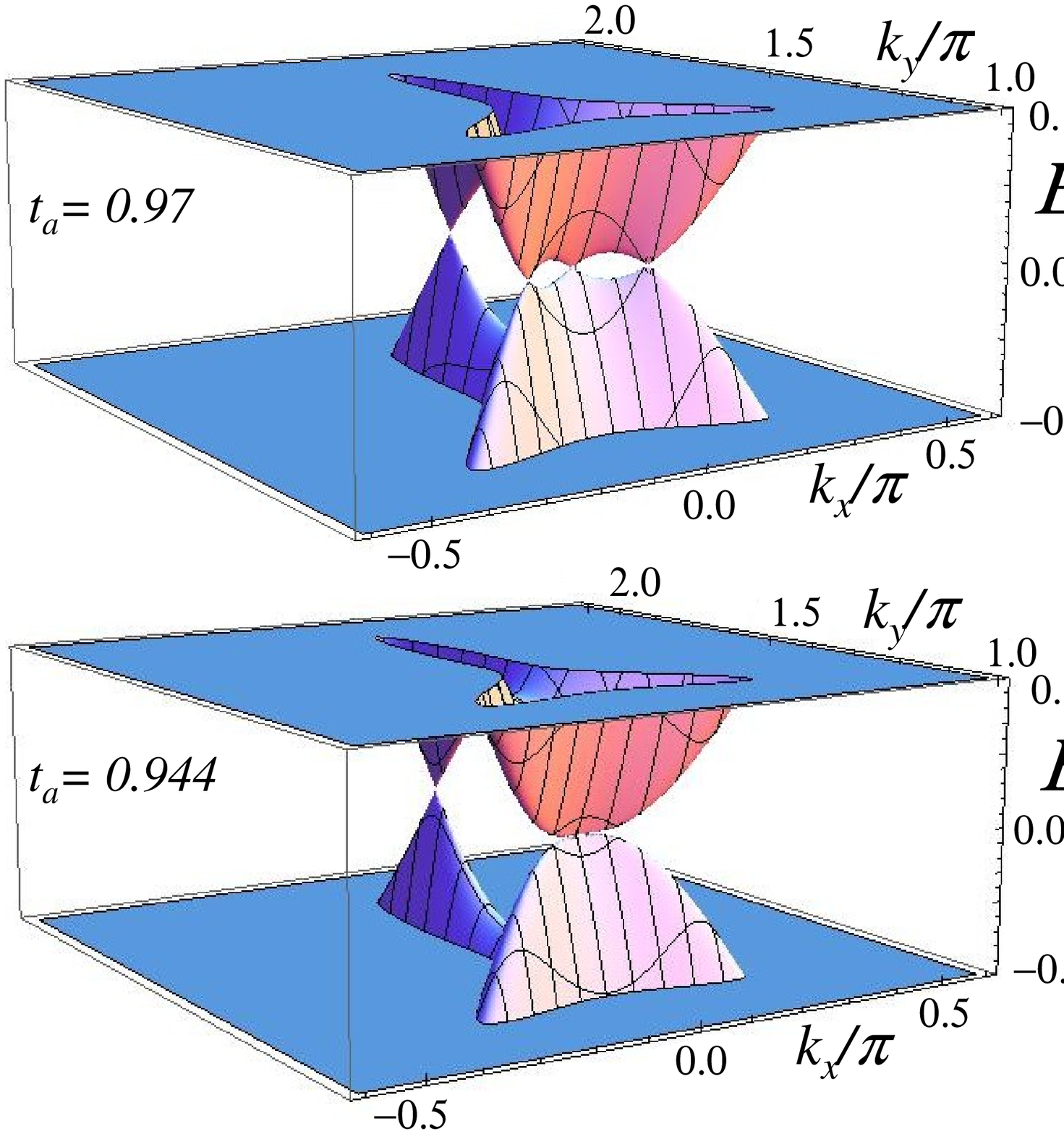}\\
\includegraphics[width=0.33\textwidth]{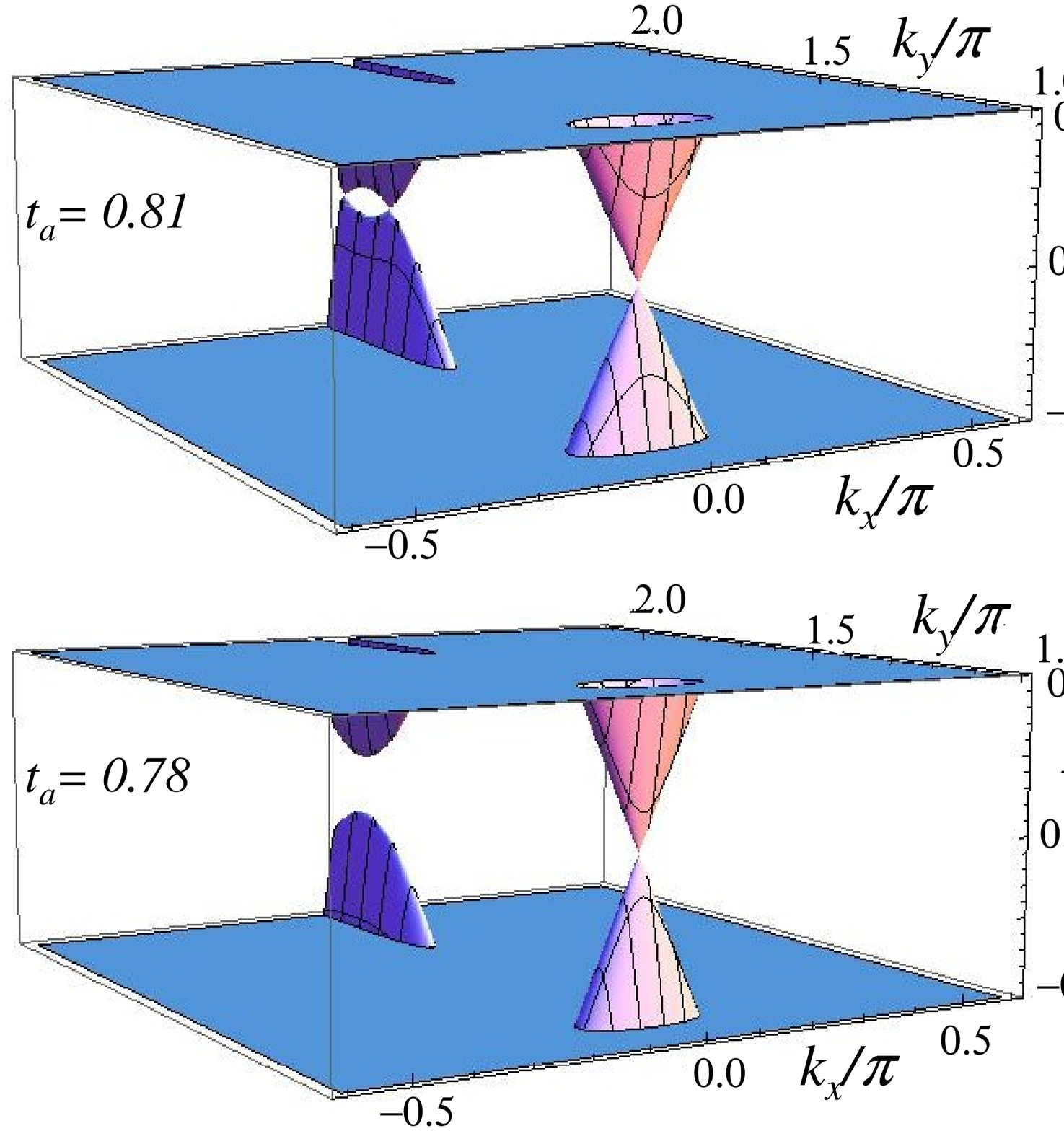}
\caption{(Color online)
3D plot of the energy as a function of the wave number
 for the parameters in Fig.\ref{fig16_e}
($t_a < 1$).
}
\label{fig16_f_2}
\end{figure}

In Fig.~\ref{fig16_e} we show how Dirac points move when $t_a$
is changed when other parameters are taken as $t_b=t_c=1$ and 
$t_{3a}=t_{3b}=t_{3c}=0.4$. 
The 3D plots of the energy are given in  Fig.\ref{fig16_f}
and Fig.\ref{fig16_f_2}.
When $t_a=1$, Dirac points are
locates at $\mathbf{K}$ and apexes of the regular triangle
(black circles in Fig.~\ref{fig16_e}. See also Fig.~\ref{figmath04}.).
As $t_a$ becomes larger, two Dirac points on the $k_y$ axis
comes closer and merge when $t_a=41/40=1.025$ 
(this set of parameters corresponds to $\mathbf{Q}_2$ 
or $\mathbf{Q}_3$ in Fig.~\ref{fig16_gx}~(b)
with exchanging $t_a$ and $t_b$ (or $t_c$)). These Dirac points disappear 
when $t_a>1.025$. Other Dirac points move to $\mathbf{M}_1$ and $\mathbf{M}_3$
and they merge when $t_a=1.2$. 
When $t_a$ becomes smaller, three of the four Dirac points 
near $\mathbf{K}$ merge when
$t_a=0.944$ (this set of parameters corresponds to 
$\mathbf{P}_2$ or $\mathbf{P}_3$ in Fig.~\ref{fig16_gx}~(b),
see also Fig.~\ref{figmath04c113}) and  one Dirac points remains  
after three Dirac points merge ($t_a<0.944$).  
Another one of the four Dirac points move to 
$\mathbf{M}_2$ and merge when $t_a = 0.8$. 
\section{Conclusions}
\label{conclusions}
We study the Dirac points of the electrons on the
honeycomb lattice with up-to third-nearest neighbor hoppings.
Since the next-nearest-neighbor hoppings 
do not affect the location of the Dirac points,
we study the location of the Dirac points and the phase diagram 
in the parameter space in the direction-dependent nearest-neighbor hoppings 
($t_a$, $t_b$ and  $t_c$) and the third-nearest-neighbor hoppings 
($t_{3a}$, $t_{3b}$ and $t_{3c}$).
Dirac points are obtained as the intersection points of two kinds of lines,
which are given by $\epsilon_{1}(\mathbf{k})=0$
and $\epsilon_{2}(\mathbf{k})=0$.
Dirac points move from $\mathbf{K}$ and $\mathbf{K}'$,
where Dirac points exist if hoppings are independent of the direction. 
We obtain the trajectories of the Dirac points as the
direction-dependent hoppings are changed.  
The topological phase transitions occur when 
Dirac points merge and disappear. 
We obtain the phase diagram in the 
parameter space of the direction-dependent hoppings.
In each phase there are 0, 2, 4, 6, or 8 Dirac points,
half of which have the topological number $+1$ and the other half have
the topological number $-1$. 
The number of the Dirac points in the Brillouin zone changes by 2
when two Dirac points merge at $\mathbf{M}_1$, $\mathbf{M}_2$,
$\mathbf{M}_3$, or $\boldsymbol{\Gamma}$, and it changes by 4 when
 two pairs of Dirac points merge at other points in the Brillouin zone, 
$\pm \mathbf{k}^*$.
When parameters are at $\mathbf{P}_1$, $\mathbf{P}_2$ or $\mathbf{P}_3$,
three Dirac points with the topological number $+1$, $-1$, and  $-1$
(or $-1$, $+1$ and $+1$) merge. 
Four Dirac points with topological number
 $+1$, $-1$, $-1$, and  $-1$ ($-1$, $+1$, $+1$, and  $+1$)
 merge at  $\mathbf{K}$ ($\mathbf{K}'$)
only when $t_a=t_b=t_c=2 t_{3a}=2 t_{3b}=2 t_{3c}$.
We also obtain that there are tricritical points $\mathbf{T}_1$,
$\mathbf{T}_1'$, $\mathbf{T}_2$ and $\mathbf{T}_3$,
 where three phases with
0, 2 and 4 Dirac points terminate.
There are no parameter regions where two Dirac points with the same
topological number merge.

The density of states $D(\epsilon)$ for 
$\epsilon \approx 0$ is proportional to 
$|\epsilon|$, $\sqrt{\epsilon}$, $\epsilon^{1/3}$, $\epsilon^{1/4}$,
and constant at the Dirac point, the merged point of two Dirac points
(with topological number $\pm 1$ and $\mp 1$),
the merged point of three Dirac points
(two of them have the topological number $\pm 1$ and the other has 
the topological number $\mp 1$), the tricritical point, and 
the merged point of four Dirac points
(three of then have the topological number $\pm 1$ and the other has
the topological number $\mp 1$),
 respectively. 

In order to realize these topological phase transitions,
only small third-nearest-neighbor hoppings are necessary, if 
the anisotropy is sufficiently large. 
The third-nearest-neighbor hoppings in
monolayer graphene are estimated to be 
finite (about 0.1 $t$)\cite{Reich2002}.
On the other hand,
if there are large third-nearest-neighbor
hoppings, small anisotropy is enough to realize 
these topological phase transitions.
The strained bilayer graphene is considered to have 
large effective third-nearest-neighbor 
hoppings\cite{Montambaux2012}.

Therefore, these  topological phase transitions can be realized
in the ultracold atoms on the optical lattice, 
strained monolayer graphene or strained bilayer graphene.


\appendix
%
\section{Dirac points in a symmetric system ($C_{6v}$)}
\label{refappA}
 We study the system with $C_{6v}$ symmetry in this Appendix, i.e.,
we set
\begin{equation}
 t \equiv t_a=t_b=t_c ,
\end{equation}
and
\begin{equation}
 t_3 \equiv t_{3a}=t_{3b}=t_{3c}.
\end{equation}
Then $\epsilon_1(\mathbf{k})$ and $\epsilon_2(\mathbf{k})$ 
(Eqs.~(\ref{eqepsilon1}) and (\ref{eqepsilon2}))
are written as
\begin{align}
\epsilon_1(\mathbf{k})
=& 
-t\left[ \cos ( \frac{\sqrt{3}}{3} k_x ) 
         +2\cos ( \frac{\sqrt{3}}{6} k_x ) \cos (\frac{1}{2}k_y)\right] 
\nonumber \\
&-t_{3}\left[ \cos ( \frac{2\sqrt{3}}{3} k_x ) 
            +2\cos ( \frac{\sqrt{3}}{3} k_x ) \cos (k_y) \right],
\label{eqepsilon100}
\end{align}
and
\begin{align}
\epsilon_2(\mathbf{k})
=& 
-t\left[ \sin ( \frac{\sqrt{3}}{3} k_x ) 
         -2\sin ( \frac{\sqrt{3}}{6} k_x ) \cos (\frac{1}{2}k_y)\right] 
\nonumber \\
&+t_{3}\left[ \sin ( \frac{2\sqrt{3}}{3} k_x ) 
            -2\sin ( \frac{\sqrt{3}}{3} k_x ) \cos (k_y) \right].
\label{eqepsilon200}
\end{align}
In Figures~\ref{figmath1}, \ref{figmath033},
\ref{figmath04}, and \ref{figmath05},
  we plot the lines for $\epsilon_1(\mathbf{k})=0$
(thin blue lines) and  for $\epsilon_2(\mathbf{k})=0$
(thick green lines). 
The massless Dirac points are given by the intersection points of 
the blue lines and the green lines.
We obtain $\epsilon_2(\mathbf{k})=0$ if $k_x=0$ 
from Eq.~(\ref{eqepsilon200}).
From the symmetry of the system, we obtain that  $\epsilon_2(\mathbf{k})=0$
on the thick green  lines in Fig.~\ref{figmath1} (a).
When $t_{3}=0$ there are no other solutions of $\epsilon_2(\mathbf{k})=0$. 

In order to find other solutions of $\epsilon_2(\mathbf{k})=0$ in the case of
$t_{3}\neq 0$, we expand $\epsilon_2(\mathbf{k})$ around $k_x=0$ and obtain
\begin{align}
 \epsilon_2(\mathbf{k})
&\approx
 \frac{\sqrt{3}}{3}\left(
 -t +t \cos \frac{k_y}{2} +2t_{3} -2 t_{3} \cos k_y \right) k_x
\nonumber \\ 
 &+ O(k_x^2).
\end{align}
The intersection points of green lines ($\epsilon_2(\mathbf{k})=0$) and
$k_y$ axis are obtained by 
the equation
\begin{equation}
  \frac{\sqrt{3}}{3}\left(
 -t +t \cos \frac{k_y^{(2)}}{2} 
 +2t_{3} -2 t_{3} \cos k_y^{(2)} \right) =0.
\end{equation}
We obtain
\begin{equation}
 \cos \frac{k_y^{(2)}}{2} = \frac{t \pm |t - 8 t_{3}|}{8 t_{3}}.
\end{equation}
This equation has the solution $k_y^{(2)}=2 n \pi$ with integers $n$, and
other solutions exist if $t_{3} > t/8$.
As seen in Fig.~\ref{figmath1}, \ref{figmath033},
\ref{figmath04}, and \ref{figmath05}, there are circular-like
 green line around 
$\mathbf{k}=0$.
Note that all $\Gamma$ points are not equivalent if we consider 
$\epsilon_1(\mathbf{k})$ and $\epsilon_2(\mathbf{k})$ separately,
while they are equivalent if we consider 
$\sqrt{(\epsilon_1(\mathbf{k}))^2+(\epsilon_2(\mathbf{k}))^2}$.

As seen in Fig.~\ref{figmath1}(b) and Fig.~\ref{figmath033}, 
if $t_{3} < t/3$,
Dirac points are only at $\mathbf{K}$ and $\mathbf{K}'$, 
since green lines and blue lines intersect only at  $\mathbf{K}$ 
and $\mathbf{K}'$.
In order to examine the number of Dirac points, we also expand 
$\epsilon_1(\mathbf{k})$ with respect to $k_x$ around $k_x=0$.
Then we obtain
\begin{align}
\epsilon_1(\mathbf{k}) &\approx  -t  \left(1+2 \cos\frac{k_y}{2}\right) 
  -t_{3}\left( 1+ 2\cos k_y \right) 
\nonumber \\
&+ O(k_x). 
\end{align}
We obtain from $\epsilon_1(\mathbf{k})=0$ at $k_x=0$ that
\begin{equation}
 \cos \frac{k_y^{(1)}}{2} = \frac{- t \pm |t - 2 t_{3}|}{4 t_{3}},
\label{eqepsilon100b}
\end{equation}
which gives the intersection points of blue lines ($\epsilon_1(\mathbf{k})=0$) 
and $k_y$ axis.
We obtain that the solution of Eq.~(\ref{eqepsilon100b})
is only $k_y^{(1)}=\pm{4\pi/3} + 4 n \pi$ ($\mathbf{K}$ 
and $\mathbf{K}'$, and their equivalent points) 
if $t_{3} < t /3$.
If  $t_{3} > t/3$, there exist other solutions of Eq.~(\ref{eqepsilon100b}),
which are the intersection points of blue ellipse and $k_y$ axis,
which we call $\mathbf{D}_1$ 
as shown in Fig.~\ref{figmath04}  for $t_{3} =0.4 t$.
Because of the symmetry the new Dirac points for $t_{3} > t /3$ 
make an equilateral triangles around  $\mathbf{K}$ and $\mathbf{K}'$, 
as shown in Fig.~\ref{figmath04}  ($\mathbf{D}_1$, 
$\mathbf{D}_2$ and $\mathbf{D}_3$).  
If $t_{3} = t /2$, four Dirac points 
merge at $\mathbf{K}$ and $\mathbf{K}'$ as shown in Fig.~\ref{figmath05}.

\section{Dirac points in an axisymmetric system ($C_{2v}$)}
\label{refappB}
In this appendix we study the system, which has $C_{2v}$ symmetry, i.e.
\begin{align}
 t_b &= t_c \equiv t_{bc},\\
t_{3b} &= t_{3c} \equiv t_{3bc}.
\end{align}
In this case the system has $\pi$ rotational symmetry 
and the reflectional symmetry with respect to the $x$-axis
and the vertical line $x=\sqrt{3}/6$.
We obtain,
\begin{align}
\epsilon_1(\mathbf{k})
=& 
-t_a \cos ( \frac{\sqrt{3}}{3} k_x ) 
         -2 t_{bc} \cos ( \frac{\sqrt{3}}{6} k_x ) \cos (\frac{1}{2}k_y) 
\nonumber \\
&-t_{3a} \cos ( \frac{2\sqrt{3}}{3} k_x ) 
 -2 t_{3bc}\cos ( \frac{\sqrt{3}}{3} k_x ) \cos (k_y),
\label{eqepsilon101}
\end{align}
and
\begin{align}
\epsilon_2(\mathbf{k})
=& 
-t_a \sin ( \frac{\sqrt{3}}{3} k_x ) 
+2 t_{bc} \sin ( \frac{\sqrt{3}}{6} k_x ) \cos (\frac{1}{2}k_y) 
\nonumber \\
&+t_{3a} \sin ( \frac{2\sqrt{3}}{3} k_x ) 
  -2 t_{3bc} \sin ( \frac{\sqrt{3}}{3} k_x ) \cos (k_y).
\label{eqepsilon201}
\end{align}
It is obtained that $\epsilon_2(\mathbf{k}) =0$ when $k_x=0$
and some of the Dirac points are on the $k_y$ axis,
as in the $C_{6v}$ case (Appendix~\ref{refappA}).
As in Appendix~\ref{refappA}, 
the intersection points of the green lines ($\epsilon_2(\mathbf{k})=0$)
and the $k_y$ axis are obtained by
\begin{equation}
  \frac{\sqrt{3}}{3}\left(
 -t_a +t_{bc} \cos \frac{k_y^{(2)}}{2} 
+2t_{3a} -2 t_{3bc} \cos k_y^{(2)} \right) =0. 
\end{equation}
We obtain 
\begin{equation}
  \cos \frac{k_y^{(2)}}{2} = \frac{t_{bc} \pm 
  \sqrt{t_{bc}^2 -16 t_a t_{3bc} 
+ 32 t_{3a} t_{3bc}
+32 t_{3bc}^2}}{8 t_{3bc}}
\label{eqepsilon2=0}
\end{equation}
The intersection points of the blue lines ($\epsilon_1(\mathbf{k})=0$)
and the $k_y$ axis are obtained by
\begin{equation}
  -t_a - 2 t_{bc} \cos \frac{k_y^{(1)}}{2} 
 - t_{3a} -2 t_{3bc} \cos k_y^{(1)} =0.
\end{equation}
From this equation we obtain
\begin{equation}
  \cos\frac{k_y^{(1)}}{2}=\frac{-t_{bc} 
 \pm \sqrt{t_{bc}^2-4 t_a t_{3bc}
 - 4t_{3a}t_{3bc}+8t_{3bc}^2}}{4 t_{3bc}}.
\label{eqepsilon1=0}
\end{equation} 
\subsection{three Dirac points merge}
\label{appendixB1}
\begin{figure}[bt]
\begin{flushleft}\hspace*{0.5cm} (a)\end{flushleft}
\begin{center}
\includegraphics[width=0.29\textwidth]{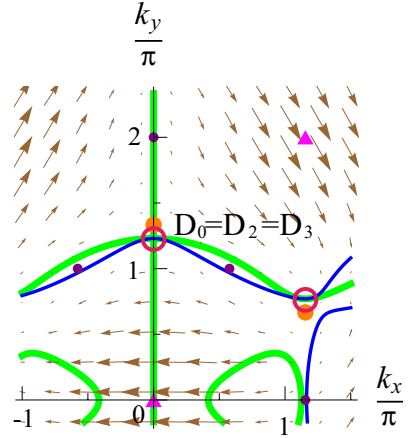}
\end{center}
\begin{flushleft}\hspace*{0.5cm} (b)\end{flushleft}
\begin{center}
\includegraphics[width=0.29\textwidth]{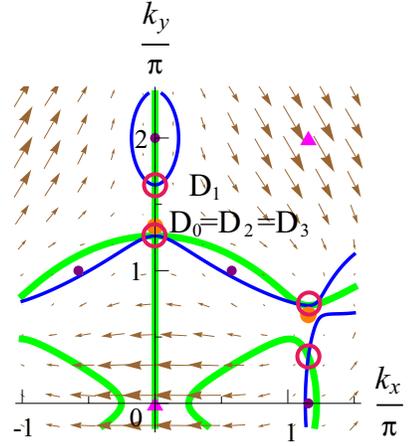}
\end{center}
\caption{(Color online)
Lines foe $\epsilon_1(\mathbf{k})=0$ (thin blue lines)
and  $\epsilon_2(\mathbf{k})=0$ (thick green lines)
for (a) $t_a=1$, $t_b=t_c=1.13137$, $t_{3a}=t_{3b}=t_{3c}=0.4$,
and
     (b) $t_a=0.944$, $t_b=t_c=1$, $t_{3a}=t_{3b}=t_{3c}=0.4$.
In both cases three Dirac points merge on the $k_y$ axis
near $\mathbf{K}$ and $\mathbf{K}'$.
The set of parameters in (a) is $\mathbf{P}_1$ in Fig.~\ref{fig16_gx}~(b),
and the set of parameters in (b) corresponds to
 $\mathbf{P}_2$ and $\mathbf{P}_3$ in Fig.~\ref{fig16_gx}~(b) 
with exchanging $t_a$ and $t_b$.
With the anisotropy of $t_a$ and $t_b=t_c$, three Dirac points 
($\mathbf{D}_0$, $\mathbf{D}_2$ and $\mathbf{D}_3$)
move from those in Fig.~\ref{figmath04} and merge.
}
\label{figmath04c113}
\end{figure}
If the solutions of Eq.~(\ref{eqepsilon2=0}) and Eq.~(\ref{eqepsilon1=0})
exist and they
are the same, three Dirac points 
($\mathbf{D}_0$, $\mathbf{D}_2$ 
and $\mathbf{D}_3$ in Fig.~\ref{figmath04})
merge at the $k_y$ axis as shown in Fig.~\ref{figmath04c113}.
We obtain that three Dirac points merge on the $k_y$ axis when
the equation
\begin{equation}
  (t_a - t_{3a} - 2 t_{3bc}) t_{bc}^2 + 4 t_{3a}^2 t_{3bc} =0
\label{eqtriple}
\end{equation}
is satisfied.
Thus, we obtain that three Dirac points merge 
on the $k_y$ axis if $(t_a, t_b(=t_{bc}), t_c (=t_{bc}))$ is given by
\begin{equation}
(t_a, t_b, t_c) = \left( t_a, 
\frac{2 t_{3a} \sqrt{t_{3bc}}}{\sqrt{-t_a + t_{3a}+ 2 t_{3bc}}},
\frac{2 t_{3a} \sqrt{t_{3bc}}}{\sqrt{-t_a + t_{3a}+ 2 t_{3bc}}}
\right),
\label{eqappa1}
\end{equation}
or 
\begin{equation}
(t_a, t_b, t_c) = \left(
  t_{3a}+2 t_{3bc}-4 \frac{t_{3a}^2 t_{3bc}}{t_{bc}^2}, 
t_{bc},t_{bc} \right).
\label{eqappa2}
\end{equation}
Eq.~(\ref{eqappa1}) is real when $t_{3a}+2 t_{3bc} > t_a$.

In Fig.~\ref{figparam4x}, Fig.~\ref{figphasenum} and Fig.~\ref{fig16_gx},
 we plot the phase diagram in the
$t_b$ - $t_c$ plane by taking 
$t_a=1$ and $t_{3a}=t_{3bc} \equiv t_{3}$.
We find three Dirac points merge when $t_b$ and $t_c$ are at
the $\mathbf{P}_1$,  $\mathbf{P}_2$, or  $\mathbf{P}_3$,
which are given 
in Eqs.~(\ref{eqeqp10}), (\ref{eqeqp20}) and (\ref{eqeqp30}).
In order to derive $\mathbf{P}_1$ we take $t_{a}=1$
in Eq.~(\ref{eqappa1}), and in order to derive
 $\mathbf{P}_2$ ($\mathbf{P}_3$)
we take $t_{bc}=1$ in Eq.~(\ref{eqappa2})
and exchange $t_a$ and $t_c$ ($t_a$ and $t_b$).
For example, if we take $t_{3a}=t_{3bc}=4/10$, we obtain three 
Dirac points merge on the $k_y$ axis if
$(t_a,t_b,t_c)=(1,4\sqrt{2}/5,4\sqrt{2}/5)\approx(1,1.13137,1.13137)$,
which corresponds to $\mathbf{P}_1$ in Fig.~\ref{fig16_gx}~(b), 
and 
$(t_a,t_b,t_c)=(0.944,1,1)$,
which corresponds to $\mathbf{P}_2$ and $\mathbf{P}_3$ 
in Fig.~\ref{fig16_gx}~(b).

In Fig.~\ref{figmath04c113} we show these two cases.
 At these parameters
three Dirac points, $\mathbf{D}_0$, $\mathbf{D}_2$ 
and $\mathbf{D}_3$, merge on the 
$k_y$ axis, as shown in Fig.~\ref{figmath04c113}. 
When parameters are 
those of Fig.~\ref{figmath04c113} (a), which correspond to $\mathbf{P}_1$ in 
Fig.~\ref{fig16_gx}~(b), $\mathbf{D}_1$ does not exist,
while  $\mathbf{D}_1$ exists when parameters are 
those of Fig.~\ref{figmath04c113} (b), which correspond to $\mathbf{P}_2$
and $\mathbf{P}_3$ in 
Fig.~\ref{fig16_gx}~(b). 
This difference can be understood
by noting the following fact in Fig.~\ref{fig16_gx} (b): When
parameters move from the point $t_a=t_b=t_c=1$ to $\mathbf{P}_1$, 
they intersect the line $M_2$,
which shows that Dirac points merge and annihilate 
at $\mathbf{M}_2$ in the Brillouin zone 
when they cross the line $M_2$.
On the other hand parameters
 do not intersect any lines labeled $M_1$, $M_2$
and $M_3$ when they move from the point $t_a=t_b=t_c=1$ to 
$\mathbf{P}_2$ or $\mathbf{P}_3$.

When parameters are changed through 
these critical values, three Dirac points merges and 
one Dirac point survives.
The topological number (Berry phase) of the Dirac point at
$\mathbf{D}_0$ is $+1$, while these at $\mathbf{D}_1$, 
$\mathbf{D}_2$ and $\mathbf{D}_3$ are $-1$.
As a result, the merged point of three Dirac points has
 the topological number $-1$.
However, the merged point is not a simple Dirac point.
The energy changes linear in one direction in the wave number
and cubic in other direction, 
and the density of states is proportional to $\epsilon^{1/3}$, 
as we show in Appendix  \ref{appendixD}.
\subsection{two Dirac points merge}
\label{appendixB2}
\begin{figure}[bt]
\begin{center}
\includegraphics[width=0.29\textwidth]{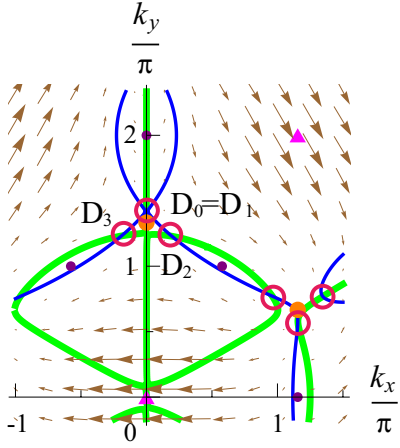}
\end{center}
\caption{(Color online)
Lines foe $\epsilon_1(\mathbf{k})=0$ (thin blue lines)
and  $\epsilon_2(\mathbf{k})=0$ (thick green lines)
for 
 $t_a=1$, $t_b=t_c=0.980$ and $t_{3a}=t_{3b}=t_{3c}=0.4$.
This set of parameters is  $\mathbf{Q}_1$ in Fig.~\ref{figparam4x} (c).
}
\label{figmath04c0980}
\end{figure}
Two Dirac points merge on the $k_y$ axis when
two solutions of Eq.~(\ref{eqepsilon1=0}) become the same,
which happens when
\begin{equation}
 t_{bc}^2 - 4 t_a t_{3bc} - 4 t_{3a} t_{3bc} + 8 t_{3bc}^2 =0,
\end{equation} 
and
\begin{equation}
 |t_{bc}| < 4 |t_{3bc}| .
\end{equation}
Thus, we obtain that two Dirac points merge 
on the $k_y$ axis if $(t_a, t_b(=t_{bc}), t_c (=t_{bc}))$ is given by
\begin{equation}
(t_a, t_b, t_c) = (\frac{t_b^2}{4 t_{3bc}}-t_{3a}+2t_{3bc},t_{bc},t_{bc}), 
\label{eqappb1}
\end{equation}
or
\begin{align}
&(t_a, t_b, t_c) \nonumber \\
=& \left( t_a,
 2\sqrt{(t_a+t_{3a}-2t_{3bc})t_{3bc}},
  2\sqrt{(t_a+t_{3a}-2t_{3bc})t_{3bc}} \right).
\label{eqappb2}
\end{align}

When $t_{3a}=t_{3bc} \equiv t_3$ and $t_a=1$,
we obtain the special points in the phase diagram 
in $t_b - t_c$ plane as,
\begin{align}
 \mathbf{Q}_1 &= \left( 2\sqrt{t_3 ( 1-t_3)},  2\sqrt{t_3 ( 1-t_3)}
 \right)
&\mbox{:($t_3 >1/5$)}, 
\label{eqeqq1}\\
 \mathbf{Q}_2 &= \left(  t_3+\frac{1}{4t_3}, 1 \right)
&\mbox{:($t_3 >1/4$)}.
\label{eqeqq2}
\end{align}
and
\begin{equation}
 \mathbf{Q}_3=\left( 1, t_3+\frac{1}{4t_3} \right)
\hspace{1cm} \mbox{:($t_3 >1/4$)}.
\label{eqeqq3}
\end{equation}

For example, if we take $t_{3a}=t_{3bc}=4/10$, we obtain 
two Dirac points merge at 
$(t_a,t_b,t_c)=(1,2\sqrt{6}/5,2\sqrt{6}/5) \approx (1,0.980,0.980)$,
which corresponds to $\mathbf{Q}_1$ in  Fig.~\ref{fig16_gx}~(b), and
at $(t_a,t_b,t_c)=(41/40,1,1) = (1.025,1,1)$,
which corresponds to $\mathbf{Q}_2$  and $\mathbf{Q}_3$ 
in Fig.~\ref{fig16_gx}~(b).
In these values of parameters, two Dirac points $\mathbf{D}_0$ 
and $\mathbf{D}_1$
merge on the $k_y$ axis as shown in Fig.~\ref{figmath04c0980}.

We obtain the phase diagram numerically in Figures  \ref{figphasenum} and
\ref{fig16_gx}.
For example, in Fig.~\ref{fig16_gx}~(b), the triangular-like region 
enclosed by the line,
$\mathbf{P}_1$-$\mathbf{S}_1$-$\mathbf{Q}_3$-$\mathbf{P}_2$-$\mathbf{Q}_1$-%
$\mathbf{P}_3$-$\mathbf{Q}_2$-$\mathbf{S}_2$-$\mathbf{P}_1$,
is the parameter region of $t_b$ and $t_c$ 
where the additional Dirac points exist when $t_a=1$, and 
$t_{3a}=t_{3b}=t_{3c}=0.4$.
Therefore, the apexes of the 
triangular-like region in Fig.~\ref{fig16_gx} (b)
are $\mathbf{P}_1=(1.131, 1.131)$, 
 $\mathbf{P}_2=(0.994, 1)$ and $\mathbf{P}_3=(1, 0.994)$,
 and the line passes on $\mathbf{Q}_1=(0.980, 0.980)$,
 $\mathbf{Q}_2=(1.025, 1)$ 
and $\mathbf{Q}_3=(1, 1.025)$.

\section{tricritical points}
\label{appendixC}
\begin{figure}[bt]
\begin{center}
\includegraphics[width=0.45\textwidth]{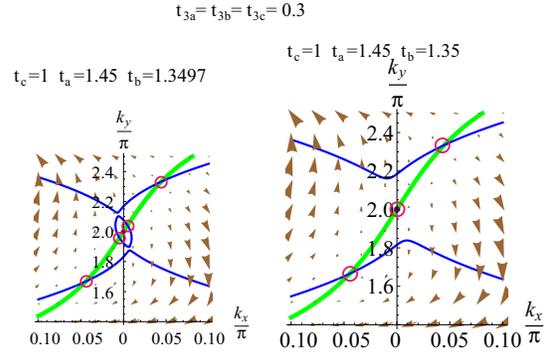}
\end{center}
\caption{(Color online)
Example for a pair of Dirac points merge at  
$\mathbf{M}$.
Lines of $\epsilon_1(\mathbf{k})=0$ (thin blue line) and 
$\epsilon_2(\mathbf{k})=0$ (thick green line).
}
\label{figt303ta145}
\end{figure}
\begin{figure}[bt]
\begin{center}
\includegraphics[width=0.45\textwidth]{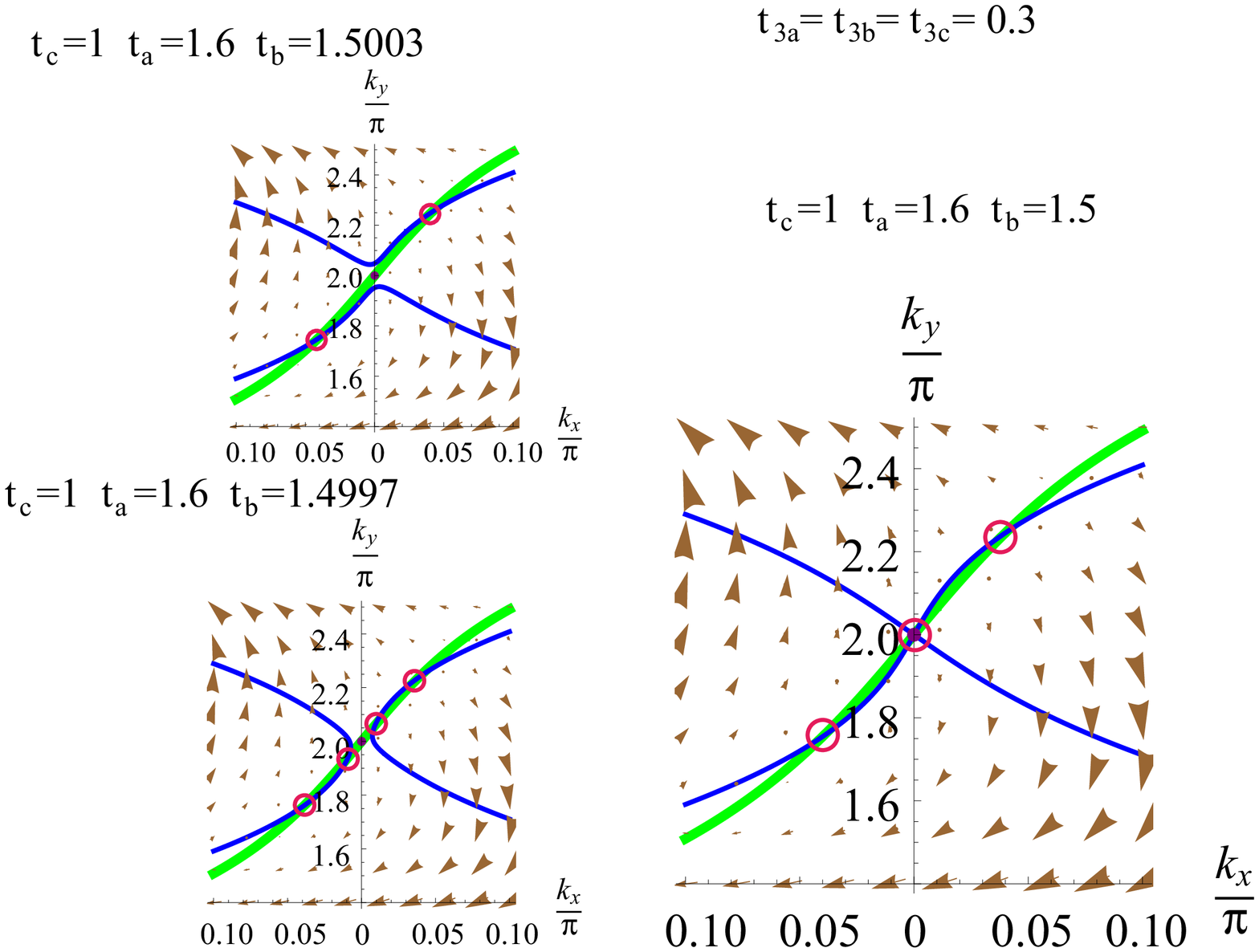}
\end{center}
\caption{(Color online)
Example for a pair of Dirac points merge at 
$\mathbf{M}$.
Lines of $\epsilon_1(\mathbf{k})=0$ (thin blue line) and 
$\epsilon_2(\mathbf{k})=0$ (thick green line).
}
\label{figt303ta16}
\end{figure}
\begin{figure}[bt]
\begin{center}
\includegraphics[width=0.45\textwidth]{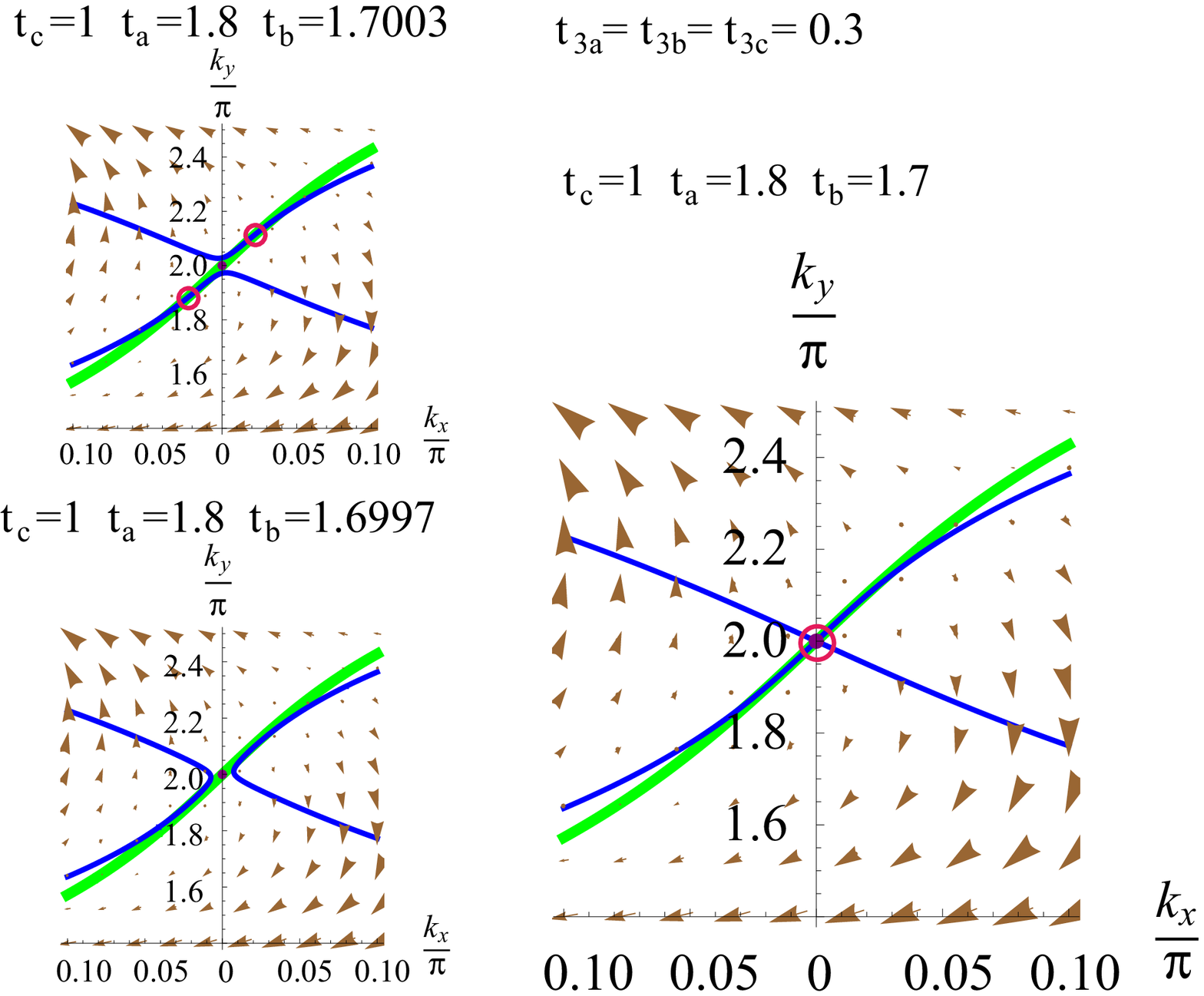}
\end{center}
\caption{(Color online)
Example for a pair of Dirac points merge at $\mathbf{M}$.
Lines of $\epsilon_1(\mathbf{k})=0$ (thin blue line) and 
$\epsilon_2(\mathbf{k})=0$ (thick green line).
}
\label{figt303ta18}
\end{figure}
In this appendix we give the analytical expression for the tricritical 
points. 
When $t_{3a}+2t_{3bc} < t_{a}$,
$\mathbf{P}_1$ does not exist
in the phase phase diagram in the $t_b/t_a$ - $t_c/t_a$ plane.
In this case the lines $\mathbf{P}_2$-$\mathbf{Q}_3$ 
and $\mathbf{P}_3$-$\mathbf{Q}_2$,
on which two Dirac points merge,
terminate tangentially on the $M_1$ and $M_3$ line 
at the tricritical points, which we call $\mathbf{T}_{3}$ 
and $\mathbf{T}_{2}$, respectively
(see Fig.~\ref{figphasenum}).
When $6t_{3bc}-t_{3a} < t_a$,
$\mathbf{Q}_1$ does not exist, and tricritical points ($\mathbf{T}_{1}$
and $\mathbf{T}_{1}'$) appear
on the $M_2$ line.
In the following we give the analytical expressions of 
$\mathbf{T}_{1}$, $\mathbf{T}_{1}'$,
$\mathbf{T}_{2}$ and $\mathbf{T}_{3}$.

We study the case
\begin{equation}
t_b =t_a-t_c+t_{3a}+t_{3b}+t_{3c}.
\end{equation}
Then merged Dirac point (semi-Dirac point)
 exists at $\mathbf{M}_2$  and the equivalent points,
for example, $\mathbf{k}=(0, 2\pi)$.
We expand $\epsilon_1(\mathbf{k})$ and $\epsilon_2(\mathbf{k})$
around $\mathbf{k}=(0, 2\pi)$ and obtain
\begin{align}
 \epsilon_1(\mathbf{k}) &\approx
 \alpha (k_y-2\pi)^2 + \beta (k_y-2\pi) k_x + \gamma k_x^2, \\
\epsilon_2(\mathbf{k}) &\approx
\delta_1 (k_y-2\pi) -\delta_2  k_x,
\end{align}
where
\begin{align}
 \alpha&=\frac{1}{8}\left(-t_a-t_{3a}+3 t_{3b}+3 t_{3c} \right),\\
 \beta&=\frac{\sqrt{3}}{12} \left( t_a-2t_c+t_{3a}-3t_{3b}+5 t_{3c} \right),
\\
 \gamma&=
 \frac{1}{8} \left( t_a+5t_{3a}+t_{3b}+t_{3c} \right), \\
\delta_1 &= \frac{1}{2} \left(t_a-2t_c+t_{3a}+3 t_{3b}-t_{3c}\right) ,
\end{align}
and
\begin{equation}
 \delta_2 = \frac{\sqrt{3}}{2} (t_a-t_{3a}+t_{3b}+t_{3c}).
\end{equation}
The line given by $\epsilon_2(\mathbf{k})=0$ goes through
$(0,2\pi)$ and it
is obtained as
\begin{equation}
 k_y-2\pi \approx \frac{\delta_2}{\delta_1} k_x.
\end{equation}
On the other hand there are no lines near $(0,2\pi)$
given by  $\epsilon_1(\mathbf{k})=0$, if 
$\beta^2-4 \alpha \gamma <0$.
We show the example in Fig.~\ref{figt303ta145}
($t_a=1.45$, $t_b=1.35$, $t_c=1$, and $t_{3a}=t_{3b}=t_{3c}=0.3$).
If $t_b$ is changed smaller,
the semi-Dirac point becomes a
a pair of Dirac points near $\mathbf{k}=(0,2\pi)$
as shown in the left figure in Fig.~\ref{figt303ta145}. 

If $\beta^2-4 \alpha \gamma >0$, 
$\epsilon_1(\mathbf{k}) =0$ gives  two lines,
\begin{equation}
   k_y-2\pi \approx 
\frac{-\beta \pm \sqrt{\beta^2-4 \alpha \gamma}}{2 \alpha} k_x,
\end{equation}
as shown in Figs.~\ref{figt303ta16} and Figs.~\ref{figt303ta18}
($(t_a,t_b,t_c)=(1.6,1.5,1)$ and $(1.8,1.7,1)$, respectively,
and  $t_{3a}=t_{3b}=t_{3c}=0.3$).

As seen in the left figures in
 Fig.~\ref{figt303ta16},
the merged Dirac points at $\mathbf{k}=(0,2\pi)$ disappear or
become two Dirac points, when we change 
$t_b$ larger or smaller, respectively.
Since  there are other Dirac points in Fig.~\ref{figt303ta16},
this topological phase transition
separate the phases with two Dirac points and four Dirac points.
On the other hand, 
as seen in the left figures in  Fig.~\ref{figt303ta18},
the merged Dirac points at $\mathbf{k}=(0,2\pi)$ 
become two Dirac points or disappear, when we change 
$t_b$ larger or smaller, respectively.
Since there are no other Dirac points in Fig.~\ref{figt303ta18},
the topological phase transition is the transition between 
two phases with two Dirac points and zero Dirac points in this parameters.
The tricritical point, on which three phases with zero, two and four 
Dirac points terminate in the parameter space, happens
when one of the lines $\epsilon_1(\mathbf{k})=0$ coincides with 
the line $\epsilon_2(\mathbf{k})=0$,
\begin{equation}
 \frac{\delta_2}{\delta_1} = 
\frac{-\beta - \sqrt{\beta^2-4 \alpha \gamma}}{2 \alpha}
\label{eqtricritical}
\end{equation} 
Tricritical points $\mathbf{T}_1$ and $\mathbf{T}_1'$ 
can be obtained by solving Eq.~(\ref{eqtricritical})
with respect to $t_c$ as a function of
$t_a$, $t_{3a}$, $t_{3b}$ and $t_{3c}$. 
We obtain 
\begin{align}
 \tilde{t}_{c\pm} &= \frac{t_a+t_{3a}+t_{3b}+t_{3c}}{2}
+ \frac{2 (t_{3b}-t_{3c})t_{3a}}{t_a+t_{3a}+t_{3b}+t_{3c}}\nonumber \\
 &\pm \frac{t_a-t_{3a}+t_{3b}+t_{3c}}{2} \nonumber \\
&\times \sqrt{\frac{t_a+t_{3a}-3(t_{3b}+t_{3c})}{t_a+t_{3a}+t_{3b}+t_{3c}}
 +\left(\frac{2(t_{3b}-t_{3c})}{t_a+t_{3a}+t_{3b}+t_{3c}}\right)^2}.
\label{eqeqtc3}
\end{align}
%
Since $\mathbf{T}_1$ and $\mathbf{T}_1'$ are on the line $M_2$
in the phase diagram in $t_b-t_c$ plane with $t_a=1$,
we obtain from Eq.~(\ref{eqeqtc3}) and taking $t_a=1$
\begin{align}
 (\mathbf{T}_{1})_x &= (1-\tilde{t}_{c-}+t_{3a}+t_{3b}+t_{3c})|_{t_a=1}
 \nonumber \\
&=\frac{1+t_{3a}+t_{3b}+t_{3c}}{2}
- \frac{2 (t_{3b}-t_{3c})t_{3a}}{1+t_{3a}+t_{3b}+t_{3c}}\nonumber \\
 &+ \frac{1-t_{3a}+t_{3b}+t_{3c}}{2} \nonumber \\
&\times \sqrt{\frac{1+t_{3a}-3(t_{3b}+t_{3c})}{1+t_{3a}+t_{3b}+t_{3c}}
 +\left(\frac{2(t_{3b}-t_{3c})}{1+t_{3a}+t_{3b}+t_{3c}}\right)^2},
\label{eqT1xA} \\
 (\mathbf{T}_{1})_y &= (\tilde{t}_{c-})|_{t_a=1}
 \nonumber \\
&=\frac{1+t_{3a}+t_{3b}+t_{3c}}{2}
+ \frac{2 (t_{3b}-t_{3c})t_{3a}}{1+t_{3a}+t_{3b}+t_{3c}}\nonumber \\
 &- \frac{1-t_{3a}+t_{3b}+t_{3c}}{2} \nonumber \\
&\times \sqrt{\frac{1+t_{3a}-3(t_{3b}+t_{3c})}{1+t_{3a}+t_{3b}+t_{3c}}
 +\left(\frac{2(t_{3b}-t_{3c})}{1+t_{3a}+t_{3b}+t_{3c}}\right)^2},
\label{eqT1yA} \\
 (\mathbf{T}_{1}')_x &= (1-\tilde{t}_{c+}+t_{3a}+t_{3b}+t_{3c})|_{t_a=1}
 \nonumber \\
&=\frac{1+t_{3a}+t_{3b}+t_{3c}}{2}
- \frac{2 (t_{3b}-t_{3c})t_{3a}}{1+t_{3a}+t_{3b}+t_{3c}}\nonumber \\
 &- \frac{1-t_{3a}+t_{3b}+t_{3c}}{2} \nonumber \\
&\times \sqrt{\frac{1+t_{3a}-3(t_{3b}+t_{3c})}{1+t_{3a}+t_{3b}+t_{3c}}
 +\left(\frac{2(t_{3b}-t_{3c})}{1+t_{3a}+t_{3b}+t_{3c}}\right)^2},
\label{eqT1pxA}
\end{align}
and
\begin{align}
 (\mathbf{T}_{1}')_y &= (\tilde{t}_{c+})|_{t_a=1}
 \nonumber \\
&=\frac{t_a+t_{3a}+t_{3b}+t_{3c}}{2}
+ \frac{2 (t_{3b}-t_{3c})t_{3a}}{t_a+t_{3a}+t_{3b}+t_{3c}}\nonumber \\
 &+ \frac{t_a-t_{3a}+t_{3b}+t_{3c}}{2} \nonumber \\
&\times \sqrt{\frac{t_a+t_{3a}-3(t_{3b}+t_{3c})}{t_a+t_{3a}+t_{3b}+t_{3c}}
 +\left(\frac{2(t_{3b}-t_{3c})}{t_a+t_{3a}+t_{3b}+t_{3c}}\right)^2}.
\label{eqT1pyA}
\end{align} 
We obtain Eqs.~(\ref{eqT1}) and (\ref{eqT1p})
 by taking $t_{3a}=t_{3b}=t_{3c}=t_3$
in Eqs.~(\ref{eqT1xA}), (\ref{eqT1yA}), (\ref{eqT1pxA}) and (\ref{eqT1pyA}).

The tricritical points $\mathbf{T}_2$ and $\mathbf{T}_3$ are obtained
as follows.
By solving 
Eq.~(\ref{eqtricritical}) with respect to $t_a$, 
we obtain the critical value of $t_a$ as a function of
$t_c$, $t_{3a}$, $t_{3b}$ and $t_{3c}$  as
\begin{align}
 \tilde{t}_a &= \frac{t_c-t_{3a}-t_{3b}-t_{3c}}{2}
+ \frac{2 (t_{3a}-t_{3b})t_{3c}}{t_c-t_{3a}-t_{3b}-t_{3c}}\nonumber \\
 &+ \frac{t_c-t_{3a}-t_{3b}+t_{3c}}{2} \nonumber \\
&\times \sqrt{\frac{t_c+3t_{3a}+3t_{3b}-t_{3c}}{t_c-t_{3a}-t_{3b}-t_{3c}}
 +\left(\frac{2(t_{3a}-t_{3b})}{t_c-t_{3a}-t_{3b}-t_{3c}}\right)^2}.
\label{eqeqta3}
\end{align} 
The tricritical point $\mathbf{T}_2$ is on the line $M_3$
in the phase diagram in $t_b-t_c$ plane,
i.e., the merged Dirac points are located at $\mathbf{M}_3$
instead of $\mathbf{M}_2$.
Since the above calculation has been done for $\mathbf{M}_2$, 
we change $t_a \to t_b \to t_c \to t_a$
and $t_{3a} \to t_{3b} \to t_{3c} \to t_{3a}$, cyclically,
in order to obtain $\mathbf{T}_2$.
 Then taking $t_a=1$,
we obtain 
 \begin{align}
 (\mathbf{T}_{2})_x &=\frac{1-t_{3a}-t_{3b}-t_{3c}}{2}
+ \frac{2 (t_{3b}-t_{3c})t_{3a}}{1-t_{3a}-t_{3b}-t_{3c}}\nonumber \\
 &+ \frac{1+t_{3a}-t_{3b}-t_{3c}}{2} \nonumber \\
&\times \sqrt{\frac{1-t_{3a}+3t_{3b}+3t_{3c}}{1-t_{3a}-t_{3b}-t_{3c}}
 +\left(\frac{2(t_{3b}-t_{3c})}{1-t_{3a}-t_{3b}-t_{3c}}\right)^2},
\label{eqT2xA}
\end{align}
and 
\begin{align}
  (\mathbf{T}_{2})_y &= (\mathbf{T}_{2})_x -1+t_{3a}+t_{3b}+t_{3c}
\nonumber \\
&=-\frac{1-t_{3a}-t_{3b}-t_{3c}}{2}
+ \frac{2 (t_{3b}-t_{3c})t_{3a}}{1-t_{3a}-t_{3b}-t_{3c}}\nonumber \\
 &+ \frac{1+t_{3a}-t_{3b}-t_{3c}}{2} \nonumber \\
&\times \sqrt{\frac{1-t_{3a}+3t_{3b}+3t_{3c}}{1-t_{3a}-t_{3b}-t_{3c}}
 +\left(\frac{2(t_{3b}-t_{3c})}{1-t_{3a}-t_{3b}-t_{3c}}\right)^2}.
\label{eqT2yA}
\end{align}
We obtain Eq.~(\ref{eqT2}) 
 by taking $t_{3a}=t_{3b}=t_{3c}=t_3$
in Eqs.~(\ref{eqT2xA}) and (\ref{eqT2yA}).

We can obtain $\mathbf{T}_3$ by 
changing $t_{3a} \to t_{3b} \to t_{3c} \to t_{3a}$, cyclically,
in Eqs.~(\ref{eqT2}) and (\ref{eqT2}).
Then we obtain Eq.~(\ref{eqT3}) similarly.

\section{Density of states}
\label{appendixD}
We study the density of states $D(\epsilon)$ for $|\epsilon| \ll 1$
 in this Appendix.
\subsection{Density of states due to Dirac points}
First we study the Density of states due to Dirac point.
A Dirac point ($\mathbf{k}^*$) 
is given by the intersection points of two lines,
 $\epsilon_1(\mathbf{k})=0$ and  $\epsilon_2(\mathbf{k})=0$,
in the Brillouin zone.
Two lines can be approximated as straight lines near the Dirac points.
We take power series expansions of $\epsilon_1(\mathbf{k})$ and  
$\epsilon_2(\mathbf{k})$ about $\mathbf{k}^*$.
By rotating the $k_x$ - $k_y$ 
axes appropriately 
and taking $\mathbf{k}^*=0$ for simplicity, we can write 
\begin{align}
 \epsilon_1(\mathbf{k}) &\approx  \frac{1}{\sqrt{2}} C  (k_x + u k_y), \\
 \epsilon_2(\mathbf{k}) &\approx  \frac{1}{\sqrt{2}} C' (k_x - u k_y),  
\end{align}
where $C (>0)$, $C' (>0)$ and $u (>0)$ are constants.
The density of states due to this Dirac point is calculated as
\begin{align}
 D(\epsilon) &=\frac{1}{S_{BZ}} \iint  d k_x d k_y 
 \delta \left( |\epsilon| -
 \sqrt{\left(\epsilon_1(\mathbf{k})\right)^2 
     + \left(\epsilon_2(\mathbf{k})\right)^2} \right)
\nonumber \\
 &= \frac{1}{S_{BZ}} \frac{ \pi}{C C' u} |\epsilon|  ,
\end{align}
where $S_{BZ}$ is the area of the Brillouin zone.
We obtain that the density of states is proportional to $|\epsilon|$.

\subsection{Density of states due to merged Dirac points}
The merged Dirac points at $\mathbf{k}=\mathbf{k}^*$
are classified into four types;
\begin{itemize}
\item[(a)] Two lines which are given by $\epsilon_1(\mathbf{k})=0$ and 
$\epsilon_2(\mathbf{k})=0$
touch in the order $n$ at $\mathbf{k}=\mathbf{k}^*$.
\item[(b)] One of the lines makes a loop and it shrinks into a point 
at $\mathbf{k}=\mathbf{k}^*$.
\item[(c)] The equation $\epsilon_2(\mathbf{k})=0$ 
gives two intersecting lines at $\mathbf{k}=\mathbf{k}^*$
and the line given by the equation $\epsilon_1(\mathbf{k})=0$
touches with one of the line in the order $m$ at $\mathbf{k}=\mathbf{k}^*$.
\item[(d)] Both equations, $\epsilon_1(\mathbf{k})=0$ and 
$\epsilon_2(\mathbf{k})=0$ give two lines intersecting at 
$\mathbf{k}=\mathbf{k}^*$.
\end{itemize}
There may be other possibilities, for example more than two
lines are given by the equation $\epsilon_2(\mathbf{k})=0$
in type (c) and (d), or two lines given by $\epsilon_2(\mathbf{k})=0$
touch each other with order $\ell (\geq 2)$ at $\mathbf{k}=\mathbf{k}^*$
in type (c).
We neglect these possibilities for simplicity, since these cases are 
not realized in our present study. 

When $t_a=t_b=t_c=1$ and $t_{3a}=t_{3b}=t_{3c}=1/3$,
two Dirac points merge at $\mathbf{M}_1$, $\mathbf{M}_2$ and $\mathbf{M}_3$
(see Fig.~\ref{figmath033}).
At 
$\mathbf{k}=(2\sqrt{3}\pi/3,0)$, two lines touch quadratically ($n=2$), 
which is type (a).
At $\mathbf{k}=(0,\pi)$, which is the equivalent point to 
$\mathbf{k}=(2\sqrt{3}\pi/3,0)$,
the line given by the equation $\epsilon_1(\mathbf{k})=0$ shrinks into a
point (compare with Fig.~\ref{figmath04}), which is type (b).
Although the merged Dirac point at $\mathbf{k}=(0,\pi)$
looks different from that at
$\mathbf{k}=(2\sqrt{3}\pi/3,0)$,
the energy around these points $\sqrt{(\epsilon_1(\mathbf{k}))^2
+(\epsilon_2(\mathbf{k}))^2}$ is the same.
Type (c) of the merged Dirac points can 
be seen in Fig.~\ref{figmath04c0980}, where
two Dirac points,
$\mathbf{D}_0$ and $\mathbf{D}_1$, merge on the $k_y$ axis when
$t_a=1$, $t_b=t_c=0.980$, and $t_{3a}=t_{3b}=t_{3c}=0.4$. 

When the parameters are at the tricritical points in the phase diagram
($\mathbf{T}_1$, $\mathbf{T}_1'$, $\mathbf{T}_2$ 
and $\mathbf{T}_3$),
 $\epsilon_1(\mathbf{k})=0$ gives two lines and 
one of the lines touches cubically  
with the line given by $\epsilon_2(\mathbf{k})=0$
(see the point $\mathbf{k}=(0, 2\pi)$ in 
Fig.~\ref{figt303ta16} and Fig.~\ref{figt303ta18}).
This is type (c) with cubic touching ($m=3$).
Four Dirac points 
with the topological numbers $+1$, $+1$, $-1$ and $-1$
merge at this point.
In the same parameters for the tricritical point, 
 type (a) with $n=4$ is realized at $\mathbf{k}=(2\sqrt{3}\pi/3,0)$,
(not shown).
  
First we study the density states for the merged Dirac points of type (a).
If two lines touch in the order $n$,
we perform the affine transformation,
 i.e. rotation, translation, and stretching in the momentum space,
and 
we can write
\begin{align}
 \epsilon_1^{(a)}(\mathbf{k}) &\approx \frac{1}{\sqrt{2}}  (k_x + u_n k_y^n), \\
 \epsilon_2^{(a)}(\mathbf{k}) &\approx \frac{1}{\sqrt{2}}  (k_x - u_n k_y^n),  
\label{eqmergetypea}
\end{align}
where  $u_n (>0)$ is a constant.
The topological number at this point is $\pm 1$ if $n$ is odd and $0$ if $n$
is even.
In this case we obtain
\begin{equation}
 \sqrt{\left( \epsilon_1^{(a)}(\mathbf{k})\right)^2
      +\left( \epsilon_2^{(a)}(\mathbf{k})\right)^2}
= \sqrt{ k_x^2 +u_n^2 k_y^{2n}}.
\end{equation}
The density of states is obtained as
\begin{equation}
D^{(a)}(\epsilon) = \frac{1}{S_{BZ}'}
\frac{2 \sqrt{\pi}\Gamma(\frac{2n+1}{2n})}
{(u_n)^{\frac{1}{n}}\Gamma(\frac{n+1}{2n})} |\epsilon|^{\frac{1}{n}},
\end{equation}
where $S_{BZ}'$ is the area of the Brillouin zone after 
the affine transformation.
We obtain the Dirac point by taking $n=1$.
When two Dirac points merge, we obtain a semi-Dirac point
(energy depend linearly on the one direction of the momentum and
quadratically in the other direction), which is obtained by taking $n=2$.
If three Dirac points merge,  $n=3$ is realized, as seen in
the merged points near $\mathbf{k}=(2\sqrt{3}\pi/3,2 \pi/3)$ in  
Fig.~\ref{figmath04c113}.
At the tricritical point in the phase diagram in the 
parameter space, 
$n=4$ is realized (at $\mathbf{k}=(2\sqrt{3}\pi/3,0)$
for the parameters in Fig.~\ref{figt303ta16} and Fig.~\ref{figt303ta18},
although that point is not shown in these figures.

 Next we study the density of states due to the
merged Dirac point of type (b). 
Type (b) of the merged Dirac points can be described 
after the affine transformation as
\begin{align}
 \epsilon_1^{(b)}(\mathbf{k}) &\approx   k_x, 
\label{eqmergetypeb1} \\
 \epsilon_2^{(b)}(\mathbf{k}) &\approx   C'' (k_x^2 + k_y^2), 
\label{eqmergetypeb2}
\end{align}
where $C'' ( > 0)$ is a constant.
In this simplification $\epsilon_1^{(b)}(\mathbf{k})=0$ gives a straight line
and $\epsilon_2^{(b)}(\mathbf{k})=0$ gives the point $\mathbf{k}=(0,0)$.
We obtain $\sqrt{(\epsilon_1^{(b)}(\mathbf{k}))^2
+(\epsilon_2^{(b)}(\mathbf{k}))^2} = |k_x| +O(k_x^2)$ in the $k_x$ direction
and $C'' k_y^2$ in the $k_y$ direction.
Therefore, Eqs.~(\ref{eqmergetypeb1}) and
(\ref{eqmergetypeb2}) describe semi-Dirac points
and the density of states is proportional to $\sqrt{\epsilon}$. Indeed, 
the density of states is calculated as
\begin{align}
 D^{(b)}(\epsilon) &\propto \int   \frac{2 |\epsilon|}{\sqrt{C''}
\sqrt{\epsilon^2-k_x^2} \sqrt{\sqrt{\epsilon^2-k_x^2}- C'' k_x^2}}
d k_x
\nonumber \\
 &=
\frac{2\sqrt{|\epsilon|}}{\sqrt{C''}} \int 
\frac{1}{\sqrt{1-x^2}\sqrt{\sqrt{1-x^2}- C'' \epsilon^2 x^2}} dx 
 \nonumber \\
 &\approx
 \frac{2 \sqrt{\pi} \Gamma (\frac{1}{4})}{\sqrt{C''}\Gamma(\frac{3}{4})}
\sqrt{|\epsilon|},
\end{align}
where integral should be performed where the integrand is real and
we have neglected higher order terms in $\epsilon$.

The merged Dirac point of type (c) is described by
\begin{align}
 \epsilon^{(c)}_1(\mathbf{k}) &\approx   k_x, \\
 \epsilon^{(c)}_2(\mathbf{k}) &\approx   k_y (k_x - u' k_y^{m}).  
\end{align}
In this case $\epsilon^{(c)}_1(\mathbf{k})=0$ gives a straight line and
 $\epsilon^{(c)}_2(\mathbf{k})=0$ gives an intersecting straight line and 
a curve touching to the first line at $\mathbf{k}=(0,0)$
 in the $m$th order (if $m=1$,
 $\epsilon^{(c)}_2(\mathbf{k})=0$ gives two intersecting lines).
We obtain $\sqrt{(\epsilon_1^{(c)}(\mathbf{k}))^2
+(\epsilon_2^{(c)}(\mathbf{k}))^2} = |k_x|$ in the $k_x$ direction
and $u' |k_y|^{m+1}$ in the $k_y$ direction.
Therefore, type (c) with $m$ is similar to type (a) with $n=m+1$.
We obtain the density of states due to this merged points as
\begin{equation}
 D(\epsilon)^{(c)} \propto |\epsilon|^{\frac{1}{m+1}}.
\end{equation} 

The simplest case of 
the type (d) is realized in the case
$t_a=t_b=t_c=1$ and $t_{3a}=t_{3b}=t_{3c}=1/2$ 
as discussed in Section \ref{sec6a}. In that case 
the topological number is two and the density of states is independent of
energy for $|\epsilon| \approx 0$.


\bibliography{diracpointsbib}
\end{document}